%% file: main.tex
\documentclass[%
 aip,
rsi,%
 amsmath,amssymb,
 reprint,%
]{revtex4-1}

\usepackage[dvipsnames]{xcolor}
\usepackage{graphicx}
\usepackage{dcolumn}
\usepackage{bm}

\usepackage{dcolumn}
\usepackage{epstopdf}
\usepackage{xcolor}
\usepackage{natbib}
\usepackage{subcaption} 
\usepackage{float}
\usepackage{tabularx}
\usepackage{mathrsfs}
\usepackage{upgreek}
\usepackage[a]{esvect}
\setcitestyle{square}
\usepackage[toc,page]{appendix}
\usepackage[font={small,normalfont}]{caption}
\usepackage[colorlinks=true, allcolors=blue]{hyperref}

\begin{document}

\title{Collimated gamma-ray beams from structured laser-irradiated targets -- \\how to increase the efficiency without increasing the laser intensity}

\author{O. Jansen}
\affiliation{Department of Mechanical and Aerospace Engineering, University of California at San Diego, La Jolla,
CA 92093}

\author{T. Wang}
\affiliation{Department of Mechanical and Aerospace Engineering, University of California at San Diego, La Jolla, CA 92093}

\affiliation{Center for Energy Research, University of California at San Diego, La Jolla, CA 92093}

\author{Z. Gong}
\affiliation{Center for High Energy Density Science, The University of Texas, Austin, TX 78712}

\author{X. Ribeyre}
\affiliation{Univ. Bordeaux-CNRS-CEA, Centre Lasers Intenses et Applications, UMR 5107, 33405 Talence, France}

\author{E. d'Humi\`eres}
\affiliation{Univ. Bordeaux-CNRS-CEA, Centre Lasers Intenses et Applications, UMR 5107, 33405 Talence, France}

\author{D. Stutman}
\affiliation{Extreme Light Infrastructure-Nuclear Physics (ELI-NP)/Horia Hulubei National Institute of Physics and Nuclear Engineering, 077125, Bucharest-Magurele, Romania}

\affiliation{Department of Physics and Astronomy, Johns Hopkins University, Baltimore, MD 21218}

\author{T. Toncian}
\affiliation{Institute for Radiation Physics, Helmholtz-Zentrum Dresden-Rossendorf e.V., 01328 Dresden, Germany}

\author{A. Arefiev}
\affiliation{Department of Mechanical and Aerospace Engineering, University of California at San Diego, La Jolla, CA 92093}

\affiliation{Center for Energy Research, University of California at San Diego, La Jolla, CA 92093}

\date{\today}

\begin{abstract}
Using three-dimensional kinetic simulations, we examine the emission of collimated gamma-ray beams from structured laser-irradiated targets with a pre-filled cylindrical channel. The channel guides the incident laser pulse, enabling generation of a slowly evolving azimuthal plasma magnetic field that serves two key functions: to enhance laser-driven electron acceleration and to induce emission of gamma-rays by the energetic electrons. Our main finding is that the conversion efficiency of the laser energy into a beam of gamma-rays ($5^{\circ}$ opening angle) can be significantly increased without increasing the laser intensity by utilizing channels with an optimal density. The conversion efficiency into multi-MeV photons increases roughly linearly with the incident laser power $P$, as we increase $P$ from 1 PW to 4 PW while keeping the laser peak intensity fixed at $5 \times 10^{22}$ W/cm$^2$. This scaling is achieved by using an optimal range of plasma densities in the channel between 10 and $20 n_{cr}$, where $n_{cr}$ is the classical cutoff density for electromagnetic waves. The corresponding number of photons scales as $P^2$. One application that directly benefits from such a strong scaling is the pair production via two-photon collisions, with the number of generated pairs increasing as $P^4$ at fixed laser intensity. 
\end{abstract}

\maketitle


\section{Introduction} \label{Sec-1}

Over the past decade, x-ray free electron lasers (XFELs)~\cite{RevModPhys.88.015006,Pellegrini_2016} have revolutionized multiple areas of science and technology by providing unprecedented sources of photons for detailed diagnostics of fast processes. For example, the European XFEL delivers over $10^{11}$ photons with energies in the range of 10 keV as a directed beam~\cite{Schneidmiller:FEL2017-MOP033}. The underlying mechanism of the XFEL operation is the photon emission by multi-GeV energetic electron beam that is periodically deflected by the magnetic field of an undulator.  

The next challenge is to develop a source of dense gamma-ray beams with photon energies in the multi-MeV range. However, upscaling the existing technology used at XFEL facilities may not be practical or feasible. Increasing the energies of emitted photons while maintaining a comparable photon yield would require a significant magnetic field increase. The desired field strength is well beyond what can be achieved using the existing technology, even with the proposed use of superconducting undulators~\cite{BAHRDT2018149}. It is then appropriate to ask whether other technological developments can be leveraged to overcome this limitation. 

One option is to use high-power high-intensity laser beams to drive extreme magnetic fields~\cite{lasinski1999particle,bulanov2010generation}. High-intensity laser pulses are capable of making a dense and otherwise opaque material transparent by heating electrons to relativistic energies. The heating increases the cutoff electron density for a given laser frequency, which is often referred to as the relativistically induced transparency\cite{palaniyappan2012dynamics,PhysRevLett.115.025002,fernandez2017laser}. This effect allows the laser pulse to volumetrically interact with a dense plasma and drive strong currents that greatly exceed the non-relativistic Alfv\'{e}nic limit of 17 kA. It has been shown using kinetic simulations that a laser-driven current can sustain magnetic fields whose strength is hundreds of kT~\cite{nakamura2010high,Stark2016PRL,0741-3335-60-5-054006, Wang_2019_MT_detection}. Experiments are being designed to measure the field inside the target using XFEL~\cite{Wang_2019_MT_detection}.

Strong quasi-static magnetic fields enable efficient generation of gamma-rays inside the laser-irradiated plasma by not only inducing the photon emissions\cite{Stark2016PRL,0741-3335-60-5-054006}, but by also enhancing the laser-driven electron acceleration~\cite{gong2018forward,gong2019strong}. Multiple promising configurations involving high-intensity lasers have been already explored in the context of laser-driven electron acceleration~\cite{liu2013generating,vranic2018extremely} and photon emission~\cite{PhysRevLett.108.195001,Ridgers2012PRL,Brady2012PRL,ridgers2013dense,ji2014radiation,ji2014_pop,blackburn2014quantum,li2015attosecond,RR2016Chang,huang2017collimated,yu2018generation,zhu2018bright,huang2019highly,QED_domian_di2012,mackenroth2011nonlinear,gonoskov2014anomalous,gonoskov2017ultrabright,chen2018gamma,magnusson2019laser}. Here our focus is on the interplay between the laser pulse and the laser-driven magnetic field in structured targets that started to be used in experiments~\cite{snyder2019relativistic, real_target_2}$^,$\footnote{J. Williams, private communications at General Atomics, 2019}. Even though the magnetic field does no work on laser-accelerated electrons, it alters how the electrons exchange their energy with the oscillating laser electric field. In a recent study, we showed that a static azimuthal magnetic field sustained by a longitudinal current can enhance the electron energy gain to the GeV level~\cite{gong2018forward}. The deflections of these electrons by the plasma magnetic field that is hundreds of kT in strength stimulate frequent emissions of gamma-rays. The resulting gamma-ray beam has two distinct lobes (see Fig. 1).

We have previously examined this method for a PW-class laser pulse with parameters projected for the Texas PW laser system~\cite{Gaul_2016}. Using 3D particle-in-cell (PIC) simulations, we found that more than 3\% of the laser energy can be converted into multi-MeV photons at a peak intensity of $5 \times 10^{22}$ W/cm$^2$ (see Ref.~\onlinecite{Stark2016PRL}). One advantage of this approach is that the multi-MeV photons are emitted as a well-directed beam by forward-moving ultra-relativistic electrons.

In this work, we investigate how the gamma-ray photon emission scales with the incident laser power in order to provide quantitative predictions for multi-PW laser facilities, such as ELI~\cite{ELI,ELI2017}, using 3D kinetic simulations. The presented study is needed because it is not immediately clear how to extrapolate the results obtained for a 1 PW laser pulse to multi-PW pulses with the same peak intensity. A gap in the current understanding becomes apparent when considering the electron acceleration by the laser pulse in the irradiated plasma. Previously published studies specify the electron energy gain for a given peak intensity or laser amplitude only~\cite{pukhov1999channel,arefiev2016beyond,gong2018forward}. The impact of the incident power is unclear and the result might even create a misleading expectation that the power of the laser pulse is irrelevant. Our 3D simulations are performed for a fixed peak laser intensity of $5 \times 10^{22}$ W/cm$^2$, with the understanding that a substantial intensity increase remains a serious technological challenge~\cite{Tiwari:19}. A peak intensity of approximately $5 \times 10^{22}$ W/cm$^2$ is for instance expected with the 10 PW ELI-NP laser, after focusing with a F/3 parabolic mirror and reflecting the beam off a single plasma mirror~\cite{URSESCU2016}.

In addition to varying the laser power, one has to consider the target geometry when optimizing the photon yield. Our previously published results indicate that a structured target with a channel serving as an optical wave-guide delivers a well-directed photon beam~\cite{Stark2016PRL,0741-3335-60-5-054006}. In contrast to that, laser propagation is unstable in dense relativistically transparent targets~\cite{huang2017relativistic}, which makes the direction of the emitted photon beam unpredictable. The optical guiding is achieved by filling the channel with a material that becomes very transparent at the peak intensity~\cite{ji2016towards}. At the same time, the bulk of the target should have a density that is at least relativistically near-critical. We perform a channel density scan for a fixed bulk density in order to optimize the gamma-ray yield. 

Our main finding is that the energy conversion efficiency into multi-MeV photons emitted into a narrow cone ($5^{\circ}$ opening angle) increases roughly linearly with the incident laser power $P$, as we increase $P$ from 1 PW to 4 PW. As a result, the emitted power by multi-MeV photons reaches 143~TW for the 4~PW case. The corresponding number of photons scales as $P^2$. This scaling is achieved by using an optimal plasma density between 10 and $20~n_{cr}$ in the channel, where $n_{cr}$ is the classical cutoff density for electromagnetic waves. In comparison to that, the conversion efficiency for an initially empty channel is much lower and it remains relatively flat with the power increase. The power increase is particularly beneficial for generation of photons with energies above 100 MeV because of the increased electron energy gain that we observe to take place.

\begin{figure}\centering
    \includegraphics[width=.4\textwidth]{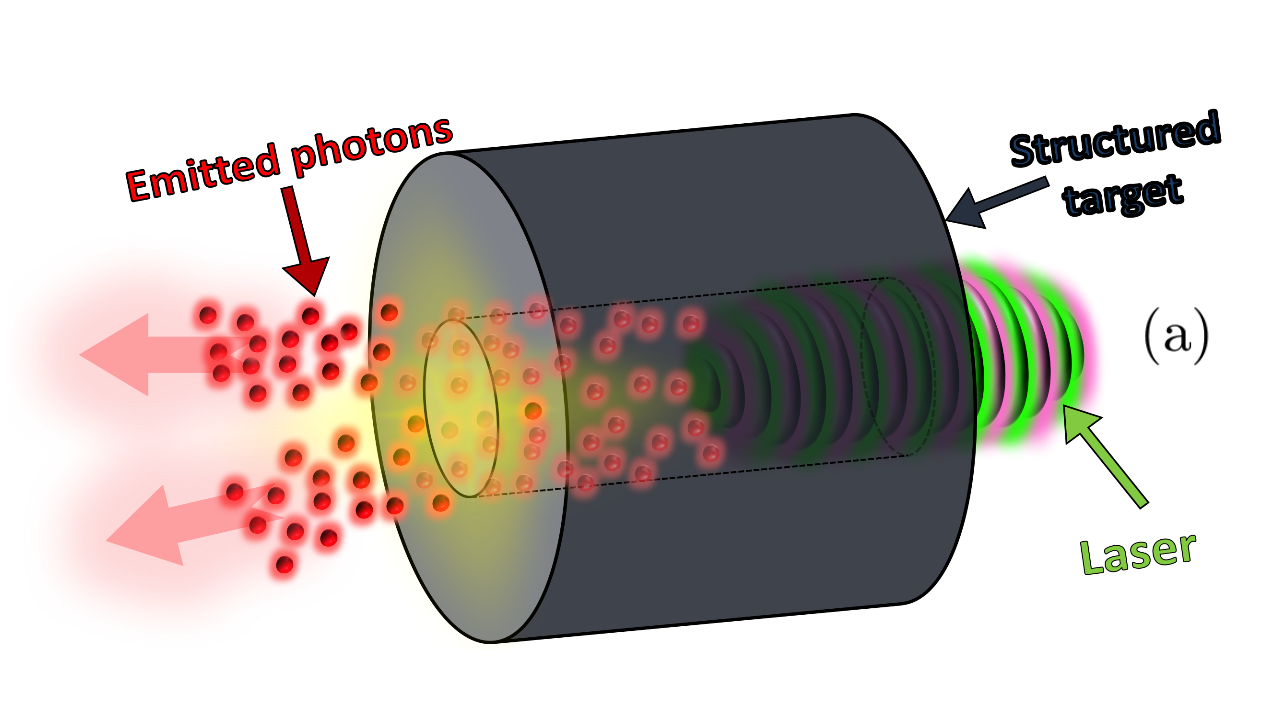}
    \includegraphics[width=.4\textwidth]{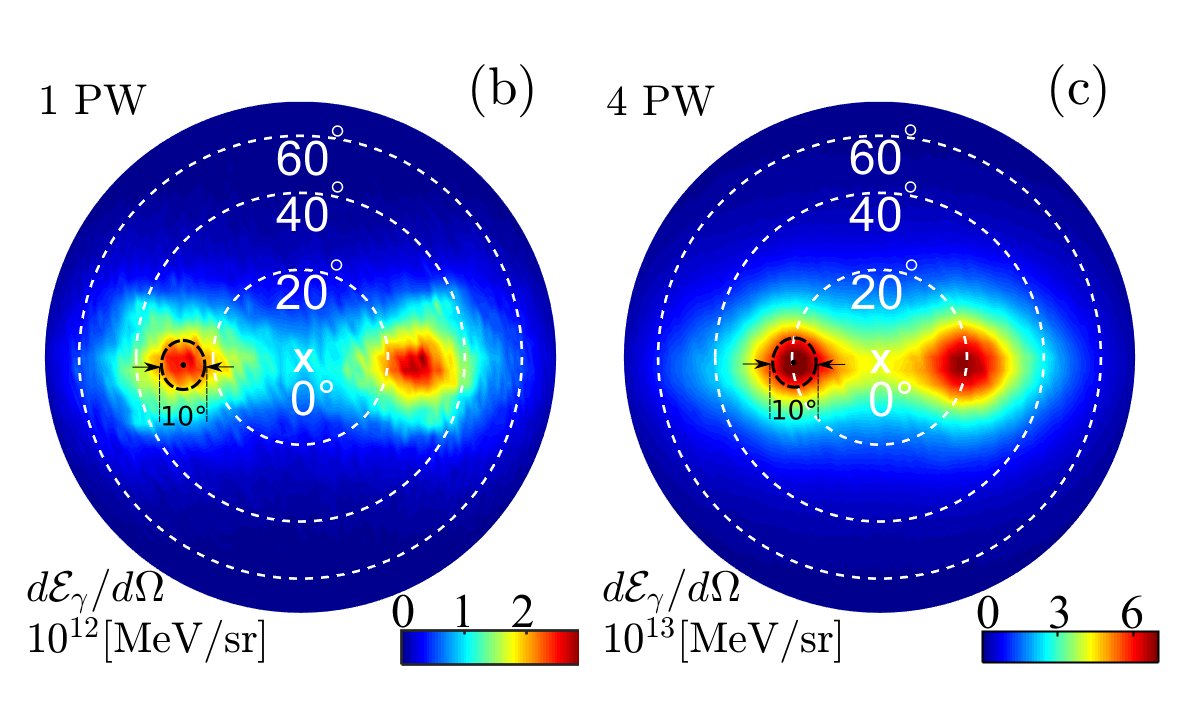}
    \includegraphics[width=.4\textwidth]{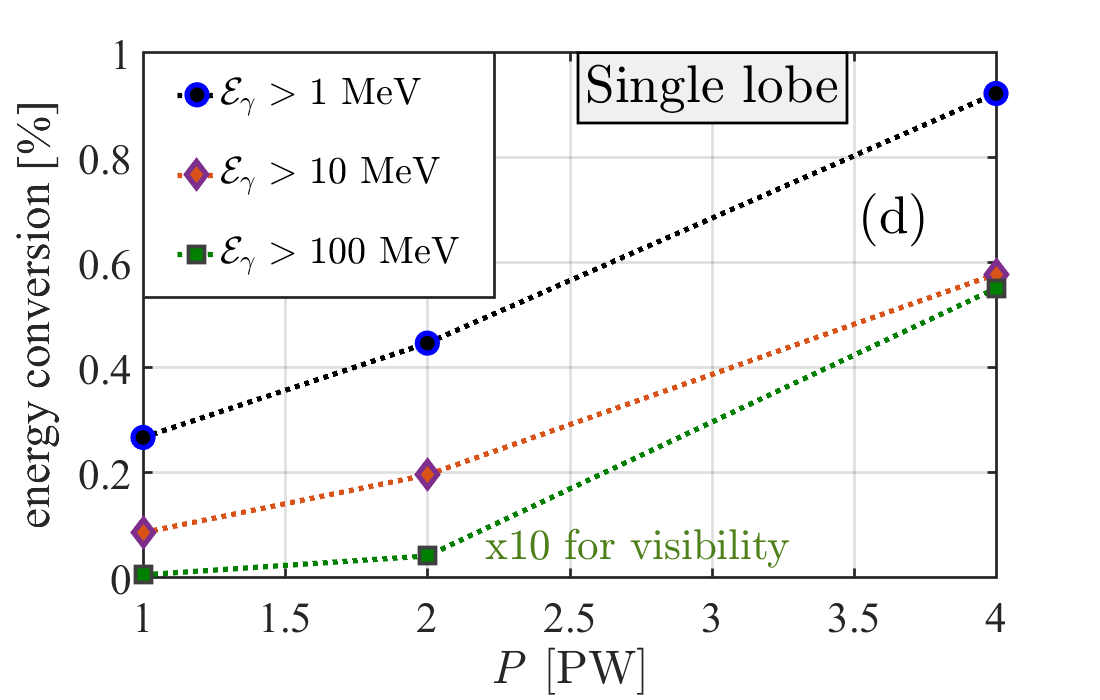}
\caption{\label{fig:SchemeLobeCon} (a) schematic setup for efficient generation of a well-directed gamma-ray beam from a laser-irradiated structured target. (b) and (c) the resulting beams of photons with $\mathcal{E}_{\gamma} >10$~MeV at incident powers of 1 and 4 PW. The dashed black circles show what we define as lobes. (d) the conversion efficiency of the laser energy into photons with energies $\mathcal{E}_\gamma$ above 1, 10, and 100 MeV emitted into a single lobe. The efficiency is shown as a function of the incident laser power $P$ for a fixed peak intensity of 5$\times 10^{22}$ W/cm$^2$.}
\end{figure}

The rest of the manuscript consists of five sections. In Section~\ref{Sec-1PW}, we discuss a baseline simulation for a 1~PW laser pulse to highlight the key features of the photon emission process and to provide the context for the discussion of multi-PW simulations. In Section~\ref{Sec-Pow}, we present results of a scan over the incident laser power $P$ for an optimal plasma density of $20~n_{cr}$ in the channel. In Section~\ref{Sec-Den}, we perform a density scan to better understand the role of the channel plasma. In Section~\ref{Sec-pair}, we apply the discussed gamma-ray sources to produce electron-positron pairs from two-photon collisions with the goal to determine how the number of the generated pairs scales with $P$. Finally, in Section~\ref{Sec-conc} we summarize our results and discuss the implications of our findings.


\section{Baseline case -- photon emission driven by a 1 PW laser pulse} \label{Sec-1PW}

Photon emission from a structured target irradiated by a PW-level laser pulse has been previously investigated and the results indicate that gamma-rays are efficiently emitted in the form of a well-directed beam~\cite{Stark2016PRL}. In this section, we review the key features of this regime that are then used in later sections of this paper to quantify and compare the performance of multi-PW laser beams.

The main points of the set-up utilizing a structured target are shown in Fig.~\ref{fig:SchemeLobeCon}a. The target consists of a cylindrical channel that is filled with a material that becomes more transparent than the bulk to the laser pulse when irradiated by a high amplitude electromagnetic wave. The channel effectively serves as an optical wave-guide to the laser pulse that is focused at the channel entrance. In our simulations, we set the channel diameter to be comparable to the focal spot of the laser in order to reduce the reflection off the bulk material, allowing for most of the laser energy to enter the target through the channel. As the laser beam propagates through the channel, it drives a longitudinal electron current that generates and sustains a quasi-static azimuthal magnetic field. The magnetic field plays two important roles: it enhances the energy gain of laser-accelerated electrons and it induces emission of gamma-rays by deflecting energetic electrons~\cite{gong2018forward, Stark2016PRL}. The synergy of these two processes within the channel leads to a significant enhancement of the gamma-ray yield compared to a regime where the interaction is restricted to the target surface and to a regime of near-vacuum interaction where collective magnetic fields play no role.

In the considered setup, the choice of the bulk and channel materials is dictated by the peak amplitude $E_0$ of the electric field in the incoming laser pulse. The laser pulse quickly ionizes the irradiated material and turns it into a plasma, so that the optical properties are determined by the electron density $n_e$ in the resulting plasma. The property that is important for our setup is the transparency. The transparency condition is defined using a dimensionless parameter that we refer to as the normalized laser amplitude: $a_0 \equiv |e| E_0 / (m_e c \omega)$, where $m_e$ and $e$ are the electron mass and charge, $\omega$ is the frequency of the laser pulse, and $c$ is the speed of light. In the case of non-relativistic plasma electrons, the plasma is transparent to the laser pulse if $n_e < n_{cr}$, where $n_{cr} \equiv m_e\omega^2/(4\pi e^2)$ is the classical critical or cutoff density. Plasma electrons become highly relativistic at $a_0 \gg 1$, which is the regime of interest for this paper. In the case of relativistic electrons with a characteristic relativistic factor $\langle \gamma \rangle$, the plasma is transparent to the laser pulse if $n_e < \langle \gamma \rangle n_{cr}$~\cite{gibbon2004short}. The increased range for the electron density has been termed the relativistically induced transparency. Typically, the value of $\langle \gamma \rangle$ resulting from electron heating by a laser pulse with a normalized amplitude $a_0$ can be estimated as $\langle \gamma \rangle \approx a_0$. Then the condition for the relativistic transparency induced by a laser pulse with $a_0 \gg 1$ is $n_e < a_0 n_{cr}$. Therefore, the plasma produced by the channel material should have $n_e \ll a_0 n_{cr}$ to enable relatively unimpeded propagation of the laser pulse. The plasma produced by the bulk material should have a much higher density to guide the laser beam. The preferred range is $n_e \sim a_0 n_{cr}$ or higher.

\begin{table}
\caption{Parameters used in the 3D PIC simulations}
\label{table_PIC}
\begin{tabular}{ |l|l| }
  \hline
  \multicolumn{2}{|c|}{3D PIC simulation parameters} \\
  \hline
  \multicolumn{2}{|l|}{\underline{\bf Laser pulse:} }\\
  Pulse energy & 35, 75, 155 J\\
  Peak intensity & $5 \times 10^{22}$ W/cm$^2$ \\
  $a_0$ & 190\\
  Wavelength & $\lambda$ = 1 $\upmu$m \\
  Power & $P$ = 1, 2 and 4 PW\\
  Location of the focal plane & $x$ = 0 $\upmu$m\\
  Pulse profile &\\
  (transverse \& longitudinal) & Gaussian \\
  Pulse duration & \\
  (FHWM for intensity) & 35 fs\\
  Pulse width/focal spot  & \\
  (FWHM for intensity) & $w_0$ = 1.3, 1.9 and 2.7 $\upmu$m\\
  \multicolumn{2}{|c|}{}\\
  \multicolumn{2}{|l|}{\underline{\bf Plasma:} }\\
  Composition & carbon ions and electrons \\
  Bulk density & 100 $n_{cr}$ \\
  Channel density & $n_{ch}$ = 0, 10, 20, and 60 $n_{cr}$ \\
  Ionization state of carbon & fully ionized \\
  Channel radius & $R_{ch}$ = 0.7 $w_0$\\
  \multicolumn{2}{|c|}{}\\
  \multicolumn{2}{|l|}{\underline{\bf General parameters:} }\\
  Spatial resolution & $30/\upmu$m $\times 30/\upmu$m $\times 30/\upmu$m \\
  $\#$ of macro-particles/cell  &  \\
  Electrons & 15 \\
  Carbon ions & 10 \\
  \hline
\end{tabular}
\end{table}


As a baseline, we have performed a 3D PIC simulation where a 1 PW laser pulse irradiates a structured target whose parameters are picked based on the discussed criteria. The parameters of the pulse are shown in Table~\ref{table_PIC}. The laser intensity has a Gaussian profile in the cross-section and, in the absence of the target, the diameter of the focal spot (full-width at half-maximum for the intensity) is $w_0 \approx 1.3 $ $\mu$m. The corresponding laser peak intensity is 5$\times 10^{22}$ W/cm$^2$. Taking into account that the vacuum wavelength is $\lambda = 1$ $\mu$m, we find that the corresponding normalized amplitude is $a_0 \approx 190$. The structured target for this laser pulse consists of a bulk plasma with $n_e = n_{bulk} \equiv 100~n_{cr}$ and a narrow channel with radius $R_{ch} \approx 0.7 w_0$ and $n_e = n_{ch} \equiv 20~n_{cr}$. The channel electron density satisfies the criterion for the relativistically induced transparency with a significant margin, $n_{ch}/n_{cr} \ll a_0$. In contrast to this, we have $n_e/n_{cr} = n_{bulk}/n_{cr} \sim a_0$ in the bulk. In line with our expectations, this allows the bulk plasma to guide the laser beam through the channel. In Sec.~\ref{Sec-Den}, it is shown that $n_{ch} = 20 n_{cr}$ provides the best photon yield and this is why we used this value here. The beam is focused at the entrance of the channel located at $x = 0$, with the $x$-axis directed into the target along the axis of the channel. The channel radius is chosen to be somewhat less than $w_0$, with $R_{ch} / w_0 \approx 0.7$. Even at $R_{ch}/w_0 \approx 0.7$ the channel entrance can capture 3/4 of the incoming power. This fraction increases during the interaction, as the channel expands appreciably. An important consideration for choosing $R_{ch} / w_0$ is the suppression of instabilities that cause the laser beam to deviate from its original direction of propagation~\cite{huang2017relativistic}. For example, we found that the propagation becomes unstable at $R_{ch} / w_0 \geq 1$.

The target material in our simulation is initialized as fully ionized plastic, represented by carbon ions (see Table~\ref{table_PIC}). No ionization takes place during the simulation, which significantly reduces computational costs during the parameter scans performed in later sections of the paper. A simulation with field ionization~\cite{Epoch} and initially neutral carbon atoms (instead of ions) reproduced the results reported in this section, which justifies our approach of using fully ionized targets.


In the described simulation with a 1 PW laser pulse, the laser pulse indeed generates high energy electrons while propagating through the channel. Figure~\ref{fig:ElecHistoLobe} that will be discussed in later sections provides a snapshot of the electron spectrum. The maximum electron energy is 400 MeV, which corresponds to $\gamma \approx 800$. This is much higher than the characteristic relativistic factor associated with electron oscillations in the laser pulse, $\langle \gamma \rangle \approx a_0 \approx 190$. The energy increase is aided by a slowly evolving magnetic field that is generated in the channel.

The quasi-static azimuthal magnetic field, whose profile is shown in Fig.~\ref{fig:PosPh}, exceeds 0.2 MT. This field bends trajectories of the energetic electrons and the resulting acceleration leads to synchrotron emission~\cite{LLclassicfields.1975}. 400 MeV electrons in a 0.2 MT magnetic field emit short wave-length electromagnetic fields representing photons with energies exceeding 1 MeV. However, it is not feasible to directly resolve this process in our simulations, because the wave-length of a 1 MeV photon is many orders of magnitude smaller than the simulation grid-size~\cite{PhysRevE.92.023305}. This difficulty is circumvented by emitting photons as individual particles. Such an approach has been successfully implemented in a PIC code EPOCH~\cite{Epoch} that we use to perform our simulations. The emissions occur according to a Monte-Carlo algorithm~\cite{duclous2010monte,sokolov_monte,elkina_monte,RIDGERS2014273,PhysRevE.92.023305} that utilizes appropriate synchrotron cross-sections.

The resulting photon emission pattern for photons with $\mathcal{E}_{\gamma} > 10$~MeV is shown in Fig.~\ref{fig:SchemeLobeCon}b. In this plot, the photons are projected onto a 3D sphere whose radius significantly exceeds the size of the emission region. The angle shown in the plot is the angle with respect to the propagation direction of the laser pulse. The horizontal plane is the polarization plane of the laser electric field. The emitted photons are primarily concentrated near this plane. This is an expected result, because the polarization plane is also the plane of electron oscillations driven by the laser electric field. An important feature is that the photon beam has two distinct lobes that are offset by a similar angle from the direction of the laser propagation. In what follows, we define a lobe more precisely as a cone with an opening angle of $5^\circ$. Each lobe is centered at the peak of emission for a given photon energy range, as seen in Figs.~\ref{fig:SchemeLobeCon}b and \ref{fig:SchemeLobeCon}c.

\begin{figure}
    \begin{center}
\includegraphics[width=1\columnwidth]{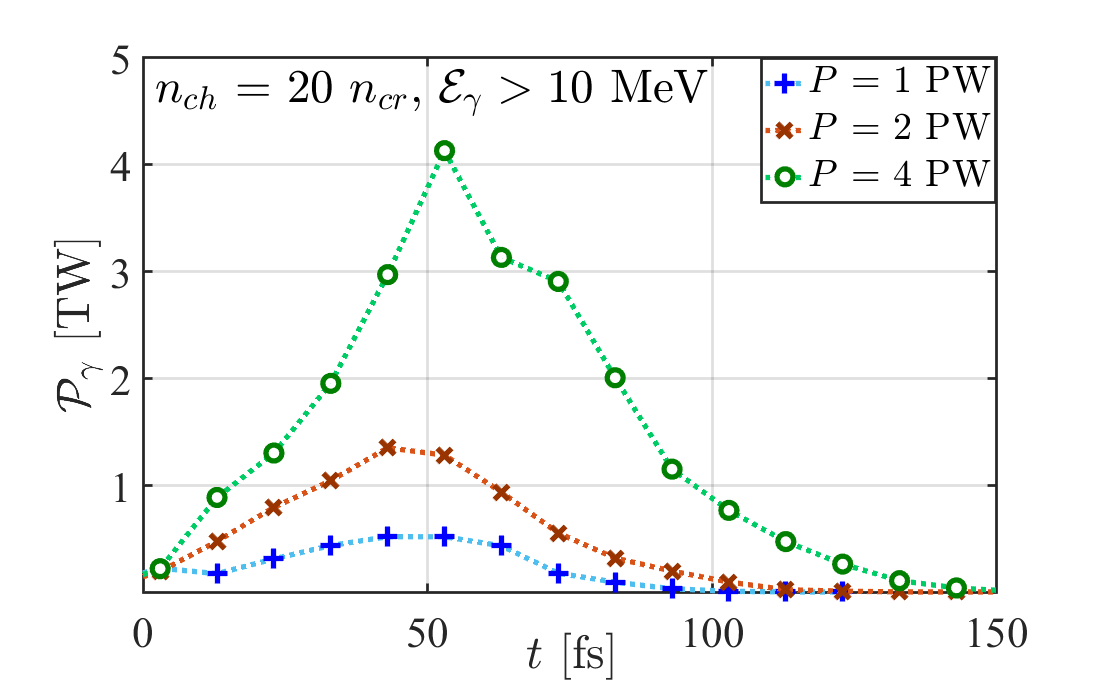}
       \caption{\label{fig:IphOvert} Normalized emitted power $\mathcal{P}_{\gamma}$ [see Eq.~(\ref{P_norm})] as a function of time $t$ at $P=1$, 2, and 4 PW. $\mathcal{P}_{\gamma}$ peaks at 43 fs for $P = 1$ and 2 PW and at 53 fs for $P = 4$~PW. The FWHMs for these curves are 50 fs, 50 fs, and 48 fs.}
       \end{center}
\end{figure}

We calculate the emission efficiency by considering only a single lobe. This approach was adopted over a standard calculation of the total energy conversion efficiency in order to provide a better metric for those applications that call for a well-collimated photon beam. The energy conversion efficiency into photons with energies above 1, 10, and 100 MeV is shown in the lower panel of Fig.~\ref{fig:SchemeLobeCon}. A single lobe produced by the 1 PW laser pulse contains 1.5$\times 10^{11}$ multi-MeV photons, which accounts for approximately 0.3\% of the total laser energy.


\section{Incident power scan} \label{Sec-Pow}

In this section we examine how the photon emission varies with the power $P$ of the incident laser pulse. We are specifically interested in a regime where the power is increased while the peak intensity is held constant. The next generation of laser systems is likely to achieve both higher power and higher peak intensity. However, the most significant improvements over the last several years have been associated with the increase in power~\cite{10.2307/27859566}, which has served as a motivation to consider a power scan at fixed intensity. 

Figure~\ref{fig:SchemeLobeCon}d shows the energy conversion efficiency into gamma-rays as a function of the incident power that is increased from 1 to 4 PW by increasing the focal spot. The channel radius is increased linearly with the diameter of the focal spot $w_0$. The efficiency is a ratio of the energy emitted by photons into a single lobe to the total energy in the incident laser pulse. Note that the two-lobe structure of the emitted beam is preserved during the power scan, as evident from Fig.~\ref{fig:SchemeLobeCon}c. The three curves in Fig.~\ref{fig:SchemeLobeCon}d correspond to photons with energies $\mathcal{E}_\gamma$ above 1, 10, and 100 MeV. The key result is that the conversion efficiency into multi-MeV photons ($\mathcal{E}_{\gamma} > 1$~MeV) increases roughly linearly with $P$, so that the number of emitted photons scales as $P^2$. In what follows, we examine the key factors contributing to the increase. 

There are three aspects that independently impact the discussed photon emission: 1) \textit{the power emitted by a single electron}, 2) \textit{the number of emitting electrons}, and 3) \textit{the duration of the emission}. It is important to point out that the number of emitting electrons has to increase faster than $P$ in order to cause an increase in the efficiency. This is because the laser pulse duration is fixed and the laser energy increases only as a result of the power increase. In order to explicitly take this aspect into account, we introduce a re-normalized emitted power
\begin{equation} \label{P_norm}
    \mathcal{P}_{\gamma} \equiv P_{lobe} \frac{1 \mbox{ PW}}{P}
\end{equation}
that is the actual emitted power into a single lobe, $P_{lobe}$, divided by a factor proportional to $P$ or, equivalently, to $w_0^2$. An increase in $\mathcal{P}_{\gamma}$ with power is then a clear indicator of increased energy conversion efficiency into photons.

\begin{figure}
      \includegraphics[width=1\columnwidth]{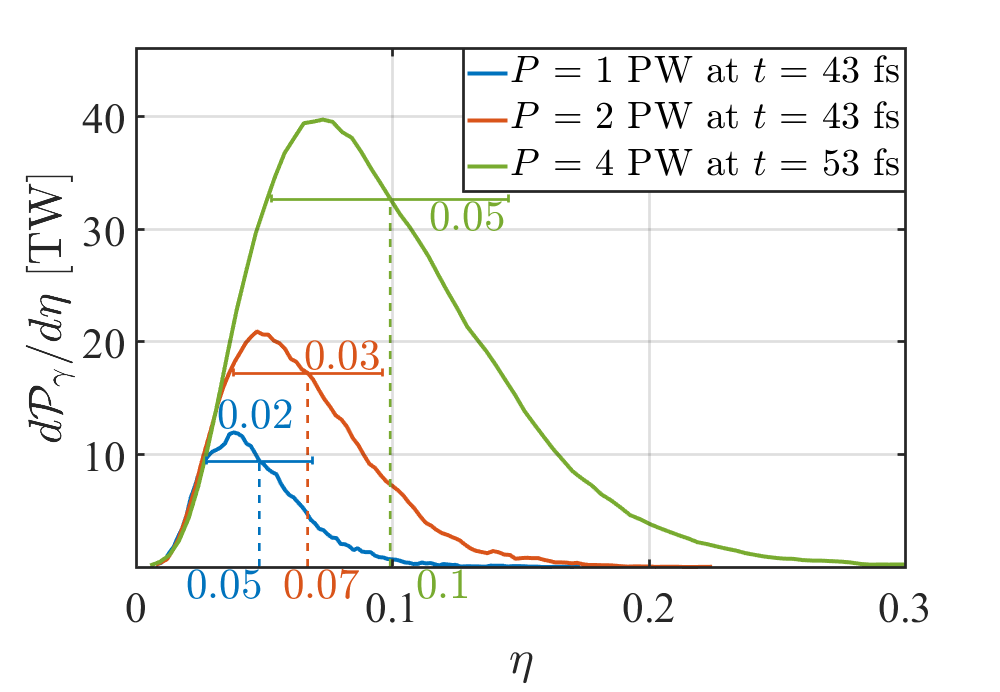}
      \caption{\label{fig:eta} Normalized emitted power $\mathcal{P}_{\gamma}$ as a function of $\eta$ at the time of maximum emission (see Fig.~\ref{fig:IphOvert}) for $P = 1$, 2 and 4 PW. $\mathcal{P}_{\gamma}$ is for photons with $\mathcal{E}_{\gamma} > 10$~MeV emitted into a single lobe. The dashed lines show the average $\eta$ and the horizontal bars show the standard deviation.}
\end{figure}

Figure~\ref{fig:IphOvert} shows the time history of $\mathcal{P}_{\gamma}$ for photons with $\mathcal{E}_{\gamma}> 10$~MeV at $P=1$, 2, and 4 PW. We define $t = 0$~fs as the time when the laser pulse reaches its peak amplitude in the focal plane at $x = 0$~$\mu$m in the absence of the target. In Figure~\ref{fig:IphOvert}, $\mathcal{P}_{\gamma}$ exhibits an increase with $P$ at the peak of the emission, which is in good agreement with the increase in the efficiency shown in Fig.~\ref{fig:SchemeLobeCon}d. The emission peaks at roughly the same time and the full width at half maximum of the curves is also roughly the same for these three values of $P$. We can thus rule out the increase in the emission duration as a possible cause for the increased conversion efficiency in our power scan. 

We find that the efficiency increase in the power scan results from increased emission by individual electrons, discussed below in Sec.~\ref{Sec-3A}, and from increased number of emitting electrons, discussed below in Sec.~\ref{Sec-3B}. We first examine the change in synchrotron emission of individual electrons. In order to be quantitative, we consider the emission of photons with $\mathcal{E}_{\gamma} > 10$ MeV into a single lobe defined in Sec.~\ref{Sec-1PW}.


\subsection{Increased emission by individual electrons} \label{Sec-3A}

The power of synchrotron emission $P_{synch}$ is determined exclusively by the electron acceleration in an instantaneous rest frame~\cite{LLclassicfields.1975}. This acceleration is proportional to a dimensionless parameter $\eta$ defined as
\begin{equation} \label{eta-AA}
    \eta \equiv \frac{\gamma_e}{E_S} \sqrt{ \left({\bf E} + \frac{1}{c}\left[{\bf v}\times{\bf B}\right]\right)^2 - \frac{1}{c^2}\left({\bf E}\cdot{\bf v}\right)^2 },
\end{equation}
where $\bf{E}$ and $\bf{B}$ are the electric and magnetic fields acting on the electron, $\gamma_e$ and $\bf{v}$ are the relativistic factor and velocity of the electron, $c$ is the speed of light, and $E_S \approx 1.3 \times 10^{18}$ V/m is the Schwinger field. The power of the synchrotron emission scales as 
\begin{equation}
    P_{synch} \propto \eta^2.
\end{equation}

\begin{figure}
      \includegraphics[width=1\columnwidth]{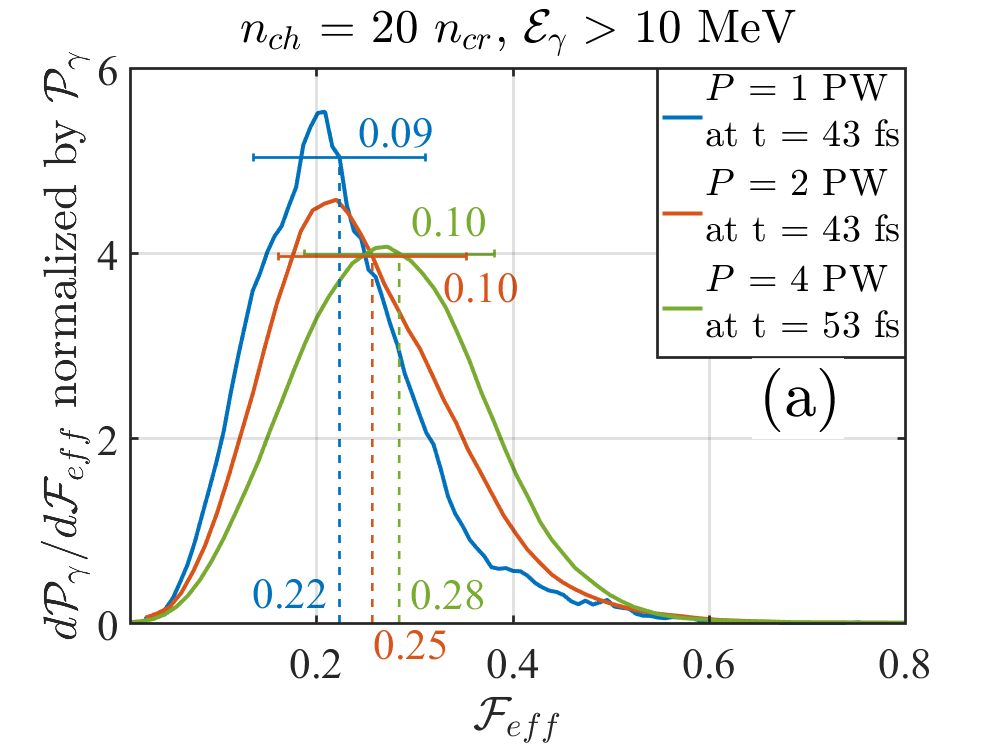}
      \includegraphics[width=1\columnwidth]{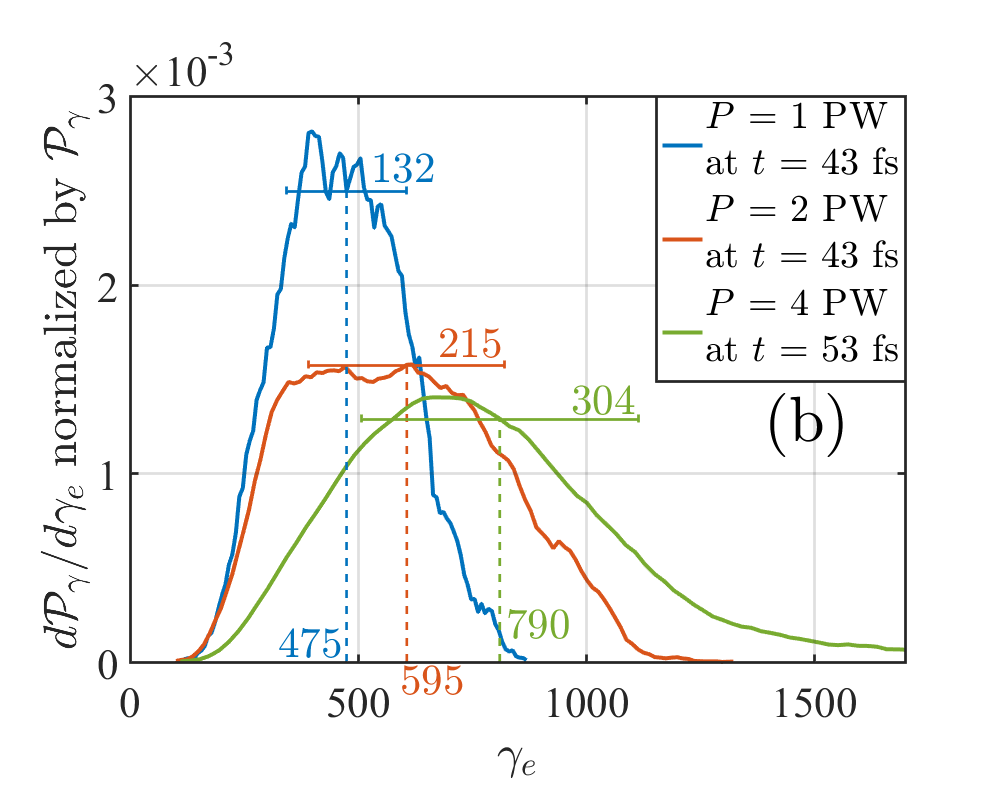}
      \caption{\label{fig:FieldGamma} Normalized emitted power $\mathcal{P}_{\gamma}$ as a function of (a) $\mathcal{F}_{eff}$ and (b) $\gamma_e$ at the time of peak emission (see Fig.~\ref{fig:IphOvert}) for $P = 1$, 2 and 4 PW. $\mathcal{P}_{\gamma}$ is for photons with $\mathcal{E}_{\gamma} > 10$~MeV emitted into a single lobe. The dashed lines show the average $\mathcal{F}_{eff}$ and $\gamma_e$. The horizontal bars show the standard deviation.}
\end{figure}

The characteristic $\eta$ during the photon emission increases with $P$, as seen in Fig.~\ref{fig:eta}. The figure shows $\mathcal{P}_\gamma$ from Fig.~\ref{fig:IphOvert} at the peak of the emission as a function of $\eta$ of the emitting electrons. Each curve has a pronounced peak. The laser power increase causes the peak to shift to the right. The average value of $\eta$ calculated for each of the curves and given in Fig.~\ref{fig:eta} also increases with $P$. The key conclusion is that an emitting electron typically emits more at higher incident laser power.

The two major factors that can increase $\eta$ are the increase in $\gamma_e$ and the change in the field configuration. The later includes not only an increase in the field strength, but also changes in the relative orientation between the electron velocity and the fields. In order to quantify this aspect, we introduce an effective field strength defined as 
\begin{equation} \label{F_eff_AA}
    \mathcal{F}_{eff} \equiv \frac{1}{E_0}\sqrt{ \left({\bf E} + \frac{1}{c}\left[{\bf v}\times{\bf B}\right]\right)^2 - \frac{1}{c^2}\left({\bf E}\cdot{\bf v}\right)^2},
\end{equation}
so that
\begin{equation} \label{eta_AA2}
    \eta = \gamma_e \mathcal{F}_{eff} E_0 / E_S,
\end{equation}
where $E_0$ is the maximum electric field strength of the laser pulse in the absence of the target, i.e. in vacuum. According to this definition, $\mathcal{F}_{eff} = 1$ for a single non-relativistic electron in vacuum with $v = 0$. We have implemented detailed tracking of emissions in EPOCH, which allows us to distinguish the changes in $\eta$ caused by changes in $\mathcal{F}_{eff}$ and $\gamma_e$ of the emitting electrons. Specifically, we record the position and momentum of each photon at the moment of the emission together with the corresponding $\eta$ and $\gamma_e$ of the emitting electron.


\subsubsection{Changes in $\mathcal{F}_{eff}$ of emitting electrons} \label{Sec-3A-1}

$\mathcal{P}_{\gamma}$ as a function of the effective field strength $\mathcal{F}_{eff}$ at the time of maximum emission is shown in Fig.~\ref{fig:FieldGamma}a for $P = 1$, 2, and 4~PW. To aid the comparison, all curves are normalized such that the area under each curve is equal to unity (we use the same approach in Fig.~\ref{fig:FieldGammaDens} of Sec.~\ref{Sec-Den}). The average values for the effective field strength range between 0.22 and 0.28. According to Eq.~(\ref{eta_AA2}), this increase in $\mathcal{F}_{eff}$ corresponds to an increases in $\eta$ by a factor of 1.3. On average, $\eta$ increases by a factor of 2, as seen in Fig.~\ref{fig:eta}, which suggests that changes in $\gamma_e$ are important.

Simple estimates for the quasi-static magnetic field suggest that we might expect a much stronger increase in $\mathcal{F}_{eff}$ compared to what we have obtained from our fully self-consistent simulations. Indeed, the maximum strength of this magnetic field $\bar{B}$ scales as $\bar{B} \propto j R_{ch}$, where $j$ is the current density driven by the laser beam in the channel cross-section and $R_{ch}$ is the channel radius. If the current density does not change significantly with power, but the channel is made wider, the maximum magnetic field inside the channel has to increase. In our power scan, the channel radius is increased as $\sqrt{P}$, which leads to $\bar{B} \propto \sqrt{P}$. We could then expect for $\mathcal{F}_{eff}$ to double as we go from 1 to 4~PW in incident power.

Our simulations confirm the increase of the maximum quasi-static magnetic field inside the channel. The detailed field structure in the channel cross-section is shown in Appendix~\ref{Appendix_B}. The values for the maximum magnetic fields $\bar{B}$ at $P = 1$, 2, and 4~PW are $0.17 B_0$, $0.23 B_0$, and $0.31 B_0$, where $B_0 = 2$~MT is the maximum magnetic field of the laser pulse in vacuum. We calculate $\bar{B}$ by averaging the magnetic field over four laser periods. 

An important aspect that is missing in our estimate is that the magnetic field limits the amplitude of transverse electron displacements across the channel. The maximum displacement scales as $1/\sqrt{j}$~\cite{gong2018forward}, which means that the electron is not able to sample the increased magnetic field in a wider channel if the current density remains roughly constant with $P$. This is a possible cause for the modest increase in $\mathcal{F}_{eff}$ experienced by the electrons in our simulations despite a more substantial increase in the magnetic field at the periphery of the channel.


\subsubsection{Changes in the $\gamma$-factor of emitting electrons} \label{Sec-3A-2}

Figure~\ref{fig:FieldGamma}b shows $\mathcal{P}_{\gamma}$ as a function of the relativistic factor $\gamma_e$ of the emitting electrons for $P = 1$, 2, and 4~PW. The average value of $\gamma_e$ increases from 475 to 790. According to Eq.~(\ref{eta_AA2}), this increase corresponds to an increases in $\eta$ by a factor of 1.7. When we combine this increase with the increase in $\mathcal{F}_{eff}$ from Fig.~\ref{fig:FieldGamma}b, we recover the factor of 2 increase in $\eta$, seen in Fig.~\ref{fig:eta}. 

The observed increase in $\gamma_e$ occurs for the same peak laser intensity and laser pulse duration, which suggests that the enhanced energy gain is related to the changes in the quasi-static magnetic field inside the channel. We have previously shown that a static azimuthal magnetic field can significantly increase the energy gain by laser-accelerated electrons~\cite{gong2018forward}. This mechanism relies on radial electron confinement. The transverse displacement does depend on the transverse electron momentum and it has to be less than the channel radius for the electrons not to become lost while being accelerated. As we increase the channel radius in our power scan, we improve the electron confinement, as confirmed by particle tracking, which then allows the electrons to reach higher energies at 4 PW than at 1 PW.  


\subsection{Increased number of emitting electrons} \label{Sec-3B}

The increase in the energy conversion efficiency results not only from the increased emission by individual electrons, but also from an increase in the number of emitting electrons. In Sec.~\ref{Sec-3A}, we showed that the characteristic $\eta$ for the emission of photons with $\mathcal{E}_{\gamma} > 10$~MeV doubles as the incident power increases from 1 to 4~PW. We can then expect an increase in the conversion efficiency by a factor of 4. However, Fig.~\ref{fig:SchemeLobeCon} shows a more significant increase for $\mathcal{E}_{\gamma} > 10$~MeV. The calculated conversion efficiency increases roughly by a factor of 6. This additional increase is an indication that the number of emitting electrons increases with power beyond just the geometric factor.

\begin{figure}
    \includegraphics[width=1\columnwidth]{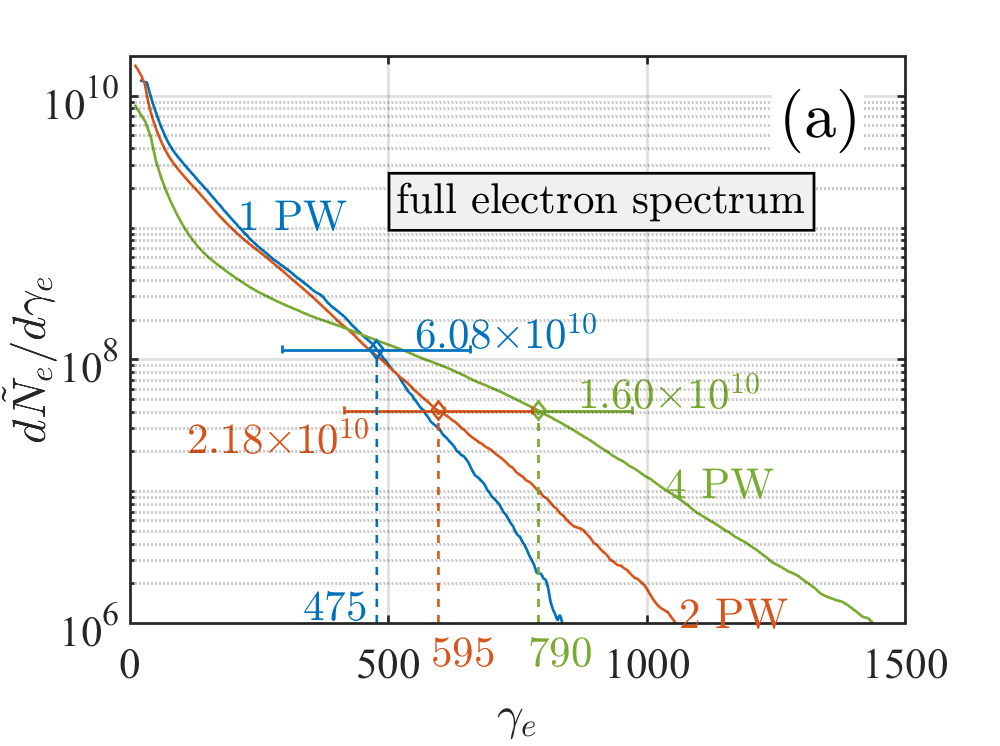}
    \includegraphics[width=1\columnwidth]{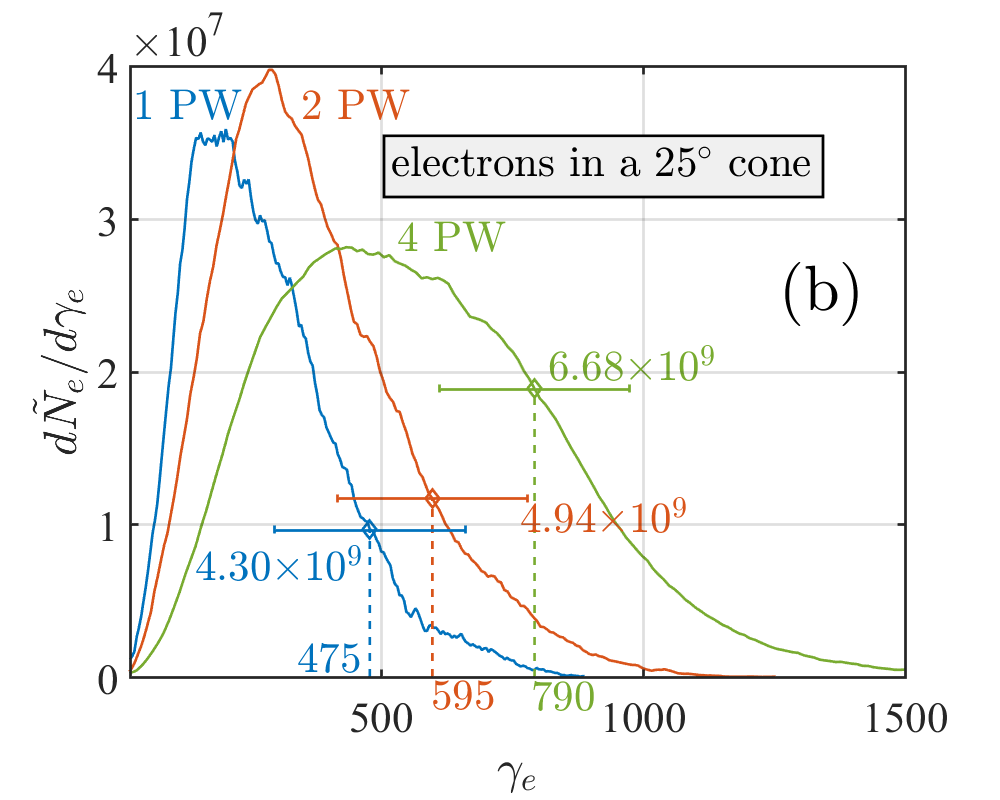}
      \caption{\label{fig:ElecHistoLobe} Normalized electron spectra at the time of peak emission (see Fig.~\ref{fig:IphOvert}) for $P = 1$, 2 and 4 PW. Only the electrons with 7 $\mu$m $\leq x \leq 12$ $\mu$m for $P = 1$ and 2 PW and with 10 $\mu$m $\leq x \leq 15$~$\mu$m for $P = 4$ PW are counted. Panel (b) shows the electrons from panel (a) that are closer than $0.5 R_{ch}$ to the axis and whose momentum is confined within a $25^{\circ}$ cone. The dashed lines show the average $\gamma_e$ from Fig.~\ref{fig:FieldGamma}b. The number next to each horizontal bar shows the electron number in the range of $\gamma_e$ represented by the bar.}
\end{figure}

Snapshots of the electron energy spectra at the time of maximum emission for $P = 1$, 2 and 4~PW (see Fig.~\ref{fig:ElecHistoLobe}a) appear to contradict the conjecture about the increase in the electron numbers. These spectra are calculated for regions with strong emission shown in Fig.~\ref{fig:PosPh} of Appendix~\ref{Appendix_B}. In each case the region is 5~$\mu$m long. The number of electrons $N$ is normalized to account for the increase in the channel radius, with the normalized number defined as
\begin{equation}
	\tilde{N} \equiv N \frac{1\mbox{ PW}}{P}.
\end{equation}
The number of electrons with the average $\gamma_e$ that is shown in Fig.~\ref{fig:FieldGamma} decreases with the increase in $P$. We have also computed the number of electrons in the vicinity of the average $\gamma_e$ that is shown with the horizontal bar for each $P$. These numbers also decrease with the increase in $P$ because the distribution function becomes less steep at higher incident power.

The apparent contradiction is due to the fact that the electron spectra in Fig.~\ref{fig:ElecHistoLobe} do not account for the fact that the electrons emit only along specific segments of their trajectories. As discussed in the preceding sections, strong fields are required for strong synchrotron emission, so the electrons emit predominantly in those regions. We have already established that the electrons emit into two lobes that are within a cone whose opening angle is roughly $25^{\circ}$ for $P = 2$ and 4~PW. The photons are emitted along the electron momentum, which means that the electron momentum at the moment of the emission is also confined within a cone whose opening angle is roughly $25^{\circ}$. Therefore, one way to account for the fact that electrons only emit photons along specific segments of the trajectory is to count electrons whose momentum is directed within the described cone. The corresponding spectra are shown in Fig.~\ref{fig:ElecHistoLobe}b. We have additionally restricted the electron sample group by counting only those electrons that are closer than $0.5 R_{ch}$ to the axis of the channel. This selection rule is motivated by the emission pattern shown in Fig.~\ref{fig:PosPh} of Appendix~\ref{Appendix_B}.

The electron spectra in Fig.~\ref{fig:ElecHistoLobe}b match our expectations, as the number of electrons with the average $\gamma_e$ from Fig.~\ref{fig:FieldGamma} increases with $P$. Moreover, the number of the electrons in the vicinity of the average $\gamma_e$ now also increases with $P$. The change in the electron numbers is likely related to the electron heating mechanism in the quasi-static azimuthal magnetic field~\cite{gong2018forward}. However, a dedicated study is required to pinpoint the parameters responsible for the trend.


\section{Channel density scan} \label{Sec-Den}

\begin{figure}
      \includegraphics[width=1\columnwidth]{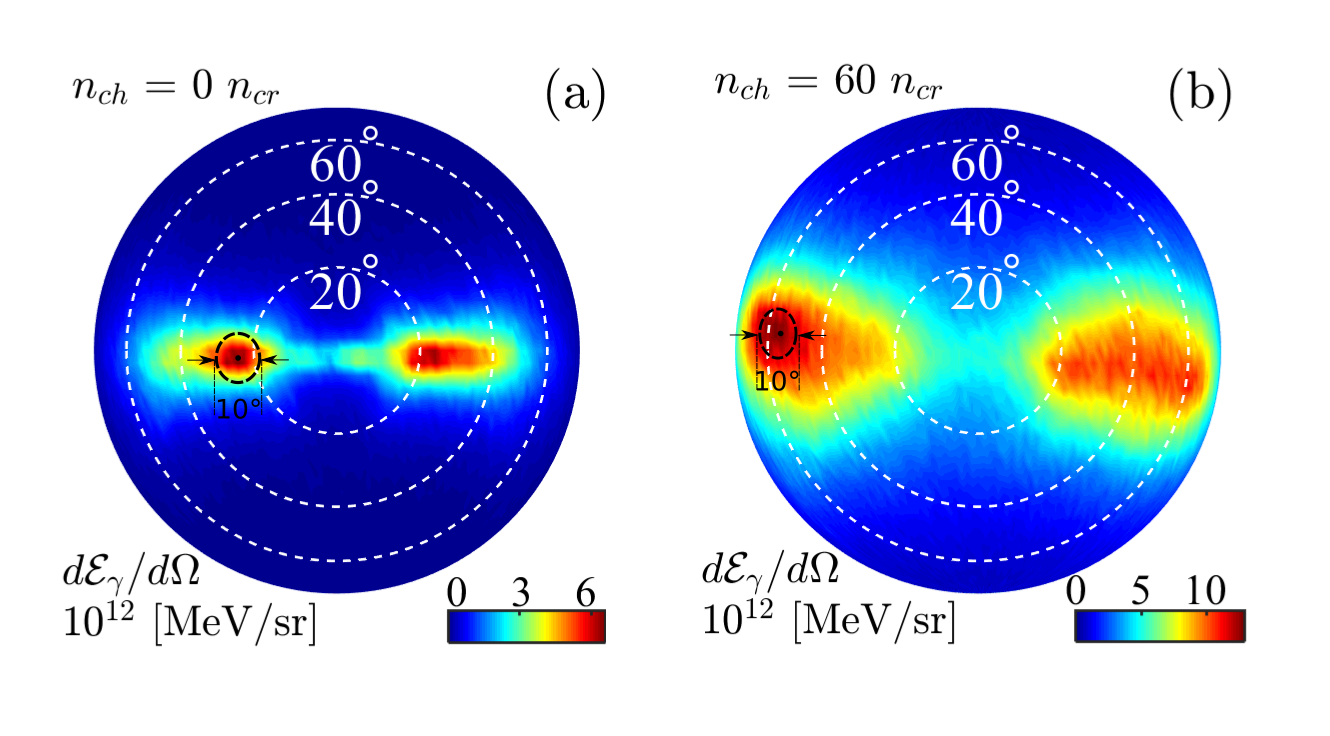}
      \caption{\label{fig:HemispheresDens} Time-integrated angular energy distribution of emitted photons with $\mathcal{E}_{\gamma} > 10$~MeV at $n_{ch} / n_{cr} = 0$ and 60 for $P$ = 4~PW. The horizontal plane is the polarization plane of the electric field in the incoming laser pulse. }
\end{figure}

In this section we perform a density scan to better understand the role of the channel plasma. In our scan we use four channel densities: $n_{ch} / n_{cr} = 0,$ 10, 20, and 60. All four values satisfy the criterion for the relativistically induced transparency, with $n_{ch} / n_{cr} < a_0 = 190$. We again only consider the photon emission into a narrow cone that is aligned with the direction of the maximum emission. We determine this direction for each set of target and laser parameters from a time-integrated angular photon emission distribution as the ones shown in Fig.~\ref{fig:HemispheresDens}.

The results of the scan are summarized in Tables~\ref{tab:Conv1}-\ref{tab:Conv100}, where the energy conversion efficiency is given for $\mathcal{E}_{\gamma} > 1$, 10, and 100~MeV. At $P = 4$~PW, $n_{ch} / n_{cr} = 10$ and 20 deliver comparable conversion rates that are significantly higher than the conversion rates for the other two densities. It is also noteworthy that the conversion rate markedly increases with incident laser power for these two densities. In contrast to that, the power increase delivers no significant increase in the conversion rate for $n_{ch} / n_{cr} = 0$ and 60 and it can even have a detrimental effect. The fact that $n_{ch} / n_{cr} = 10$ and 20 perform equally well suggests that the range of optimal densities is relatively wide and that fine-tuning of the channel density is not required to achieve the optimal performance.

In what follows, we compare the emission of photons with $\mathcal{E}_{\gamma} > 10$~MeV at $n_{ch} / n_{cr} = 0$ and 60 to the emission at $n_{ch} / n_{cr} = 20$ from the previous section. Our goal is to elucidate the underlying cause for the enhanced performance at the optimal density. We limit our consideration to $P = 4$~PW since this power delivers the highest conversion rate for the optimal channel density.


\begin{table}
\caption{Density scan for $\mathcal{E}_\gamma > 1$ MeV.}
\label{tab:Conv1}
\begin{tabular}{|l|c|c|c|c|}
\hline
conversion & $n_{ch} = 60$ $n_{cr}$ & {$n_{ch} = 20$ $n_{cr}$} & $n_{ch} = 10$ $n_{cr}$ & $n_{ch} = 0$\\
to lobe[$\%$] & & & &\\
  \hline
  \hline
  $P$ = 1 PW & 0.24 & {0.27} & 0.18 & 0.09\\
  \hline
  $P$ = 2 PW & 0.31 & {0.45} & 0.36 & 0.10\\
  \hline
  $P$ = 4 PW & 0.29 & {0.92} & 0.68 & 0.08\\
  \hline
  \end{tabular}
\bigskip
\caption{Density scan for $\mathcal{E}_\gamma > 10$ MeV.}
\label{tab:Conv10}
\begin{tabular}{|l|c|c|c|c|}
\hline
conversion & $n_{ch} = 60$ $n_{cr}$ & {$n_{ch} = 20$ $n_{cr}$} & $n_{ch} = 10$ $n_{cr}$ & $n_{ch} = 0$\\
to lobe[$\%$] & & & &\\
  \hline
  \hline
  $P$ = 1 PW & 8.5$\cdot{10^{-2}}$ & {9.6$\cdot{10^{-2}}$} & 5.8$\cdot{10^{-2}}$ & 3.5$\cdot{10^{-2}}$\\
  \hline
  $P$ = 2 PW & 14$\cdot{10^{-2}}$ & {19$\cdot{10^{-2}}$} & 18$\cdot{10^{-2}}$ & 3.1$\cdot{10^{-2}}$\\
  \hline
  $P$ = 4 PW & 15$\cdot{10^{-2}}$ & {58$\cdot{10^{-2}}$} & 40$\cdot{10^{-2}}$ & 3.0$\cdot{10^{-2}}$\\
  \hline
  \end{tabular}
\bigskip
\caption{Density scan for $\mathcal{E}_\gamma > 100$ MeV.}
\label{tab:Conv100}
\begin{tabular}{|l|c|c|c|c|}
\hline
conversion & $n_{ch} = 60$ $n_{cr}$ & $n_{ch} = 20$ $n_{cr}$ & $n_{ch} = 10$ $n_{cr}$ & $n_{ch} = 0$\\
to lobe[\%] & & & &\\
  \hline
  \hline
  $P$ = 1 PW & 2.8$\cdot{10^{-3}}$ & 3.4$\cdot{10^{-3}}$ & {4.1$\cdot{10^{-3}}$} & 1.5$\cdot{10^{-3}}$ \\
  \hline
  $P$ = 2 PW & 3.5$\cdot{10^{-3}}$ & 4.0$\cdot{10^{-3}}$ & {5.2$\cdot{10^{-3}}$} & 0.4$\cdot{10^{-3}}$ \\
  \hline
  $P$ = 4 PW & 7.5$\cdot{10^{-3}}$ & {55$\cdot{10^{-3}}$} & 31$\cdot{10^{-3}}$ & 0.7$\cdot{10^{-3}}$ \\
  \hline
  \end{tabular}
\end{table}


Figure~\ref{fig:HemispheresDens} shows qualitative differences in the angular distribution of the photons with $\mathcal{E}_{\gamma} > 10$~MeV emitted at $n_{ch} / n_{cr} = 0$ and 60. The initially empty channel produces a structure similar to that observed at the optimal density. This two-lobe structure is a clear indicator that the ion dynamics plays an important role at $n_{ch} / n_{cr} = 0$, causing transverse electron oscillations~\cite{Wang_2019}. We found that the single lobe structure reported in Ref.~[\onlinecite{yu2018generation}] can only be achieved by using a laser pulse with an extremely sharp rising edge rather than the Gaussian used here. In contrast to the regime with $n_{ch} / n_{cr} = 0$, the emission at $n_{ch} / n_{cr} = 60$ no longer exhibits strong angular localization. In this case, the emission is spread over more than $50^{\circ}$ in the polarization plane of the electric field.

\begin{figure}
      \includegraphics[width=1\columnwidth]{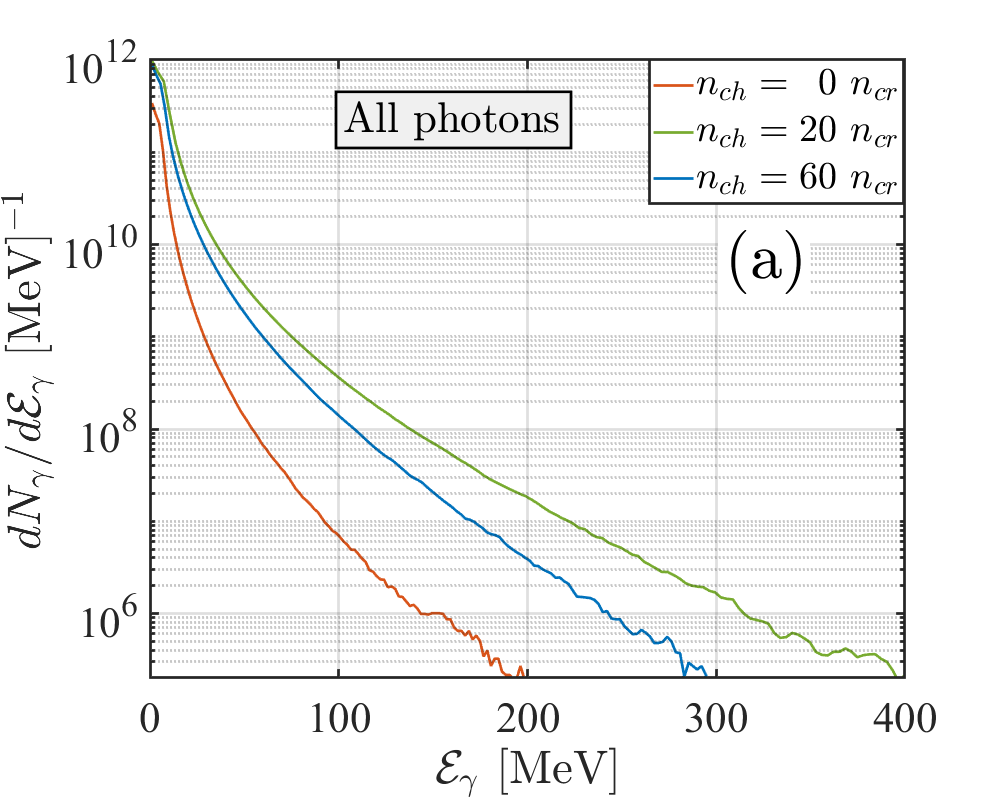}
      \includegraphics[width=1\columnwidth]{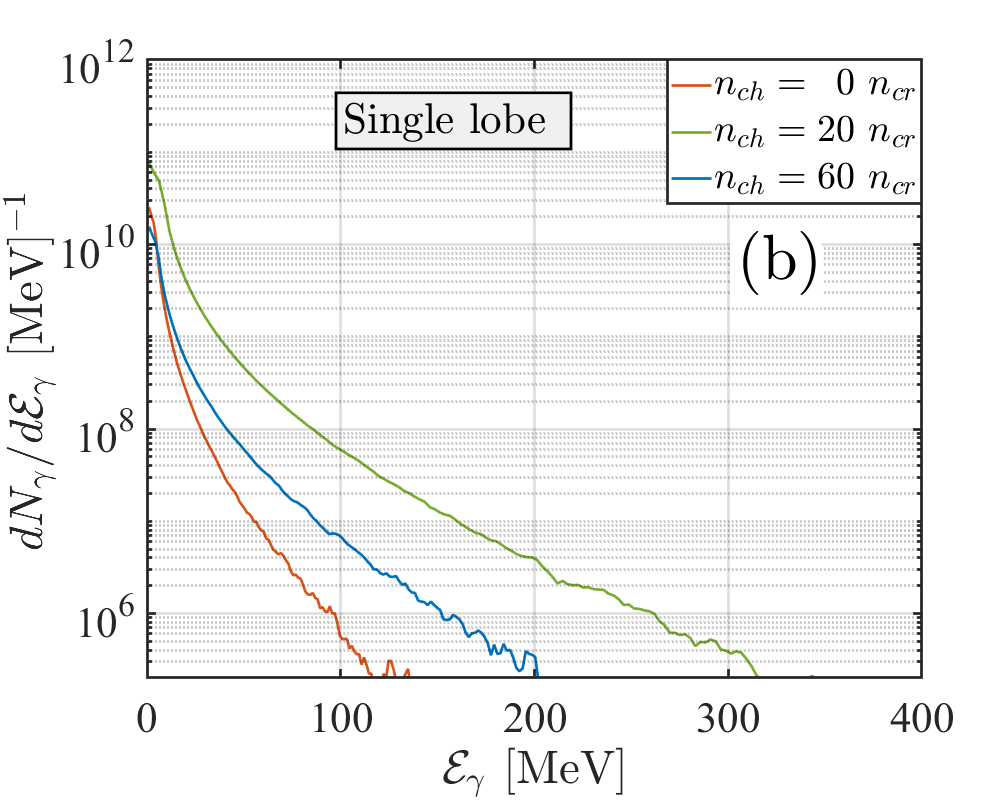}
      \caption{\label{fig:PhotEspectrumDens} Time-integrated photon spectra at $P$ = 4~PW for $n_{ch} / n_{cr} = 0,$ 20, and 60. (a) the total photon spectra and (b) the spectra of the photons emitted into a single lobe.}
\end{figure}

The wide angular spread at $n_{ch} / n_{cr} = 60$ is the cause for the reduced conversion efficiency when considering only the emission of photons into a narrow opening cone. In fact, the full photon spectra at $n_{ch} / n_{cr} = 60$ and at the optimal density of $n_{ch} / n_{cr} = 20$ are very similar for $\mathcal{E}_{\gamma} < 100$, as seen in Fig.~\ref{fig:PhotEspectrumDens}a. However, the photons are much better collimated at $n_{ch} / n_{cr} = 20$. As a result, the spectrum of photons emitted into a narrow cone aligned with the direction of maximum emission is several times higher at $n_{ch} / n_{cr} = 20$, as shown in Fig.~\ref{fig:PhotEspectrumDens}b. The angular spread of the emitted photons at $n_{ch} / n_{cr} = 60$ is a direct manifestation of the fact that the laser is unable to generate a directed beam of energetic electrons. We can thus conclude that the channel density that is much higher than the optimal density negatively impacts the laser-driven electron acceleration in the channel.

\begin{figure}
      \includegraphics[width=1\columnwidth]{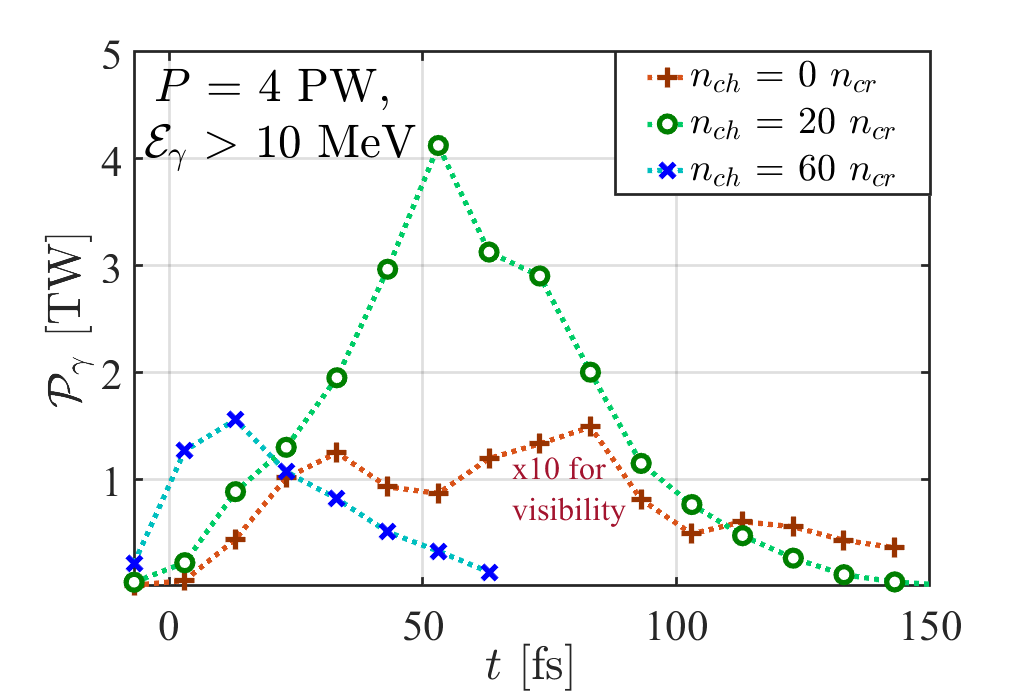}
      \caption{\label{fig:NphDens} Re-normalized emitted power $\mathcal{P}_{\gamma}$ [see Eq.~(\ref{P_norm})] at $P=4$ PW for $n_{ch} / n_{cr} = 0$, 20, and 60. $\mathcal{P}_{\gamma}$ peaks at 83 fs, 53 fs, and 13 fs, with the FWHM of 77 fs, 48 fs, and 36 fs.}
\end{figure}

In contrast to the case of $n_{ch} / n_{cr} = 60$, an initially empty channel is able to effectively accelerate electrons injected into the channel by the laser pulse and generate a directed beam of energetic electrons. We find that the laser depletion is a lot slower than at the optimal density of $n_{ch} / n_{cr} = 20$. As a consequence, the photon emission lasts longer and it peaks at a later time, as shown in Fig.~\ref{fig:NphDens}. However, the downside of using an initially empty channel is that the number of laser-injected electrons that undergo acceleration in the channel and emit photons is much less than at the optimal density.

\begin{figure}
      \includegraphics[width=1\columnwidth]{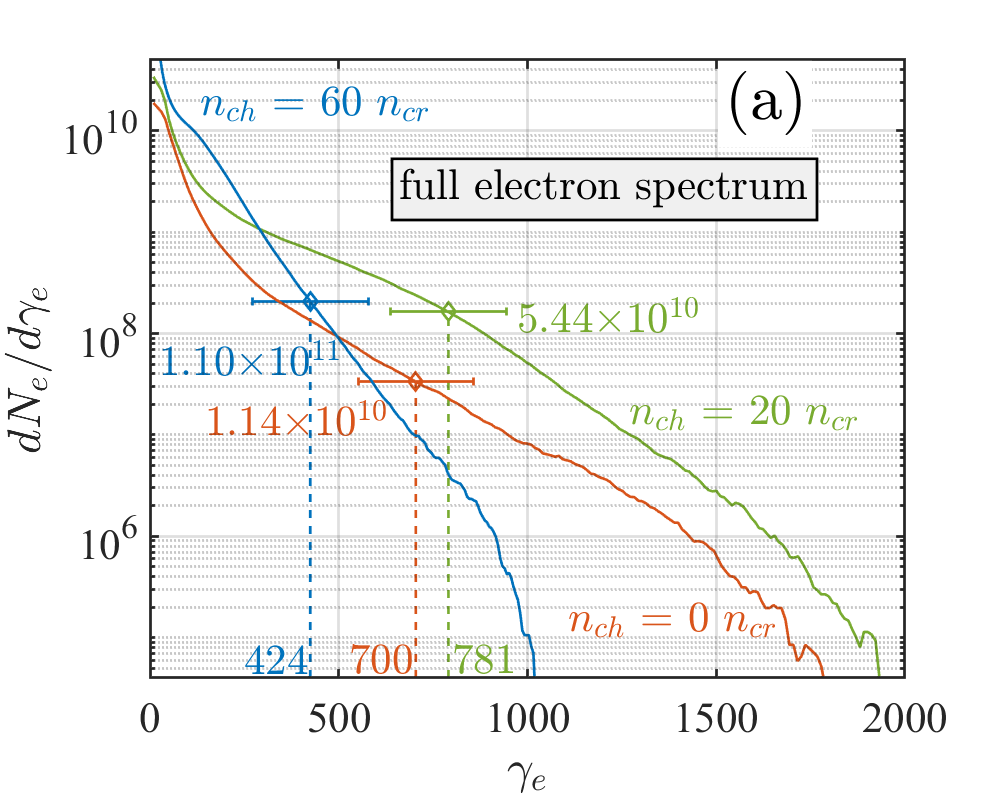}
      \includegraphics[width=1\columnwidth]{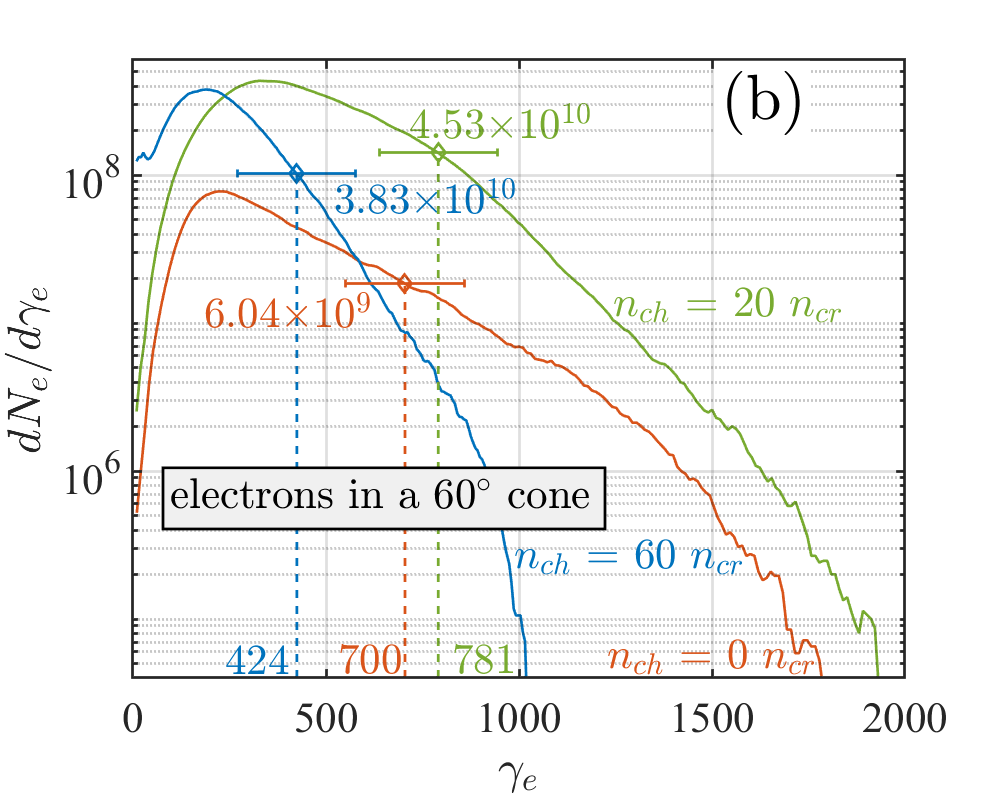}
      \caption{\label{fig:eHistoDens} Snapshots of electron spectra at the time of maximum emission from Fig.~\ref{fig:NphDens} for $n_{ch} / n_{cr} = 0$, 20, and 60. Only the electrons with 20 $\mu$m $\leq x \leq 25$ $\mu$m, 10 $\mu$m $\leq x \leq 15$~$\mu$m, and 0 $\mu$m $\leq x \leq 5$~$\mu$m are counted for these densities. Panel (b) shows a subset of electrons from panel (a) that are closer than $0.5 R_{ch}$ to the axis and whose momentum is confined within a $60^{\circ}$ cone.}
\end{figure}

Figure~\ref{fig:eHistoDens}a shows the electron spectra inside the channel at the time of peak emission from Fig.~\ref{fig:NphDens}. In each case, we only count the electrons that are located within a 5~$\mu$m vicinity of the axial location with the maximum emission. The electron spectra for $n_{ch} / n_{cr} = 0$ and 20 are very similar: they have similar slopes at $\gamma_e > 300$ and they have similar maximum electron energies. The main difference is the vertical scale, which represents the electron numbers. The number of energetic electrons at $n_{ch} / n_{cr} = 0$ is almost an order of magnitude lower. It must be pointed out that the spectrum at $n_{ch} / n_{cr} = 60$ has a much steeper slope while the maximum electron energy only reaches 500~MeV. This further substantiates our earlier statement that the laser-driven electron acceleration is negatively impacted at this channel density.

\begin{figure}
   \begin{center}
      \includegraphics[width=1\columnwidth]{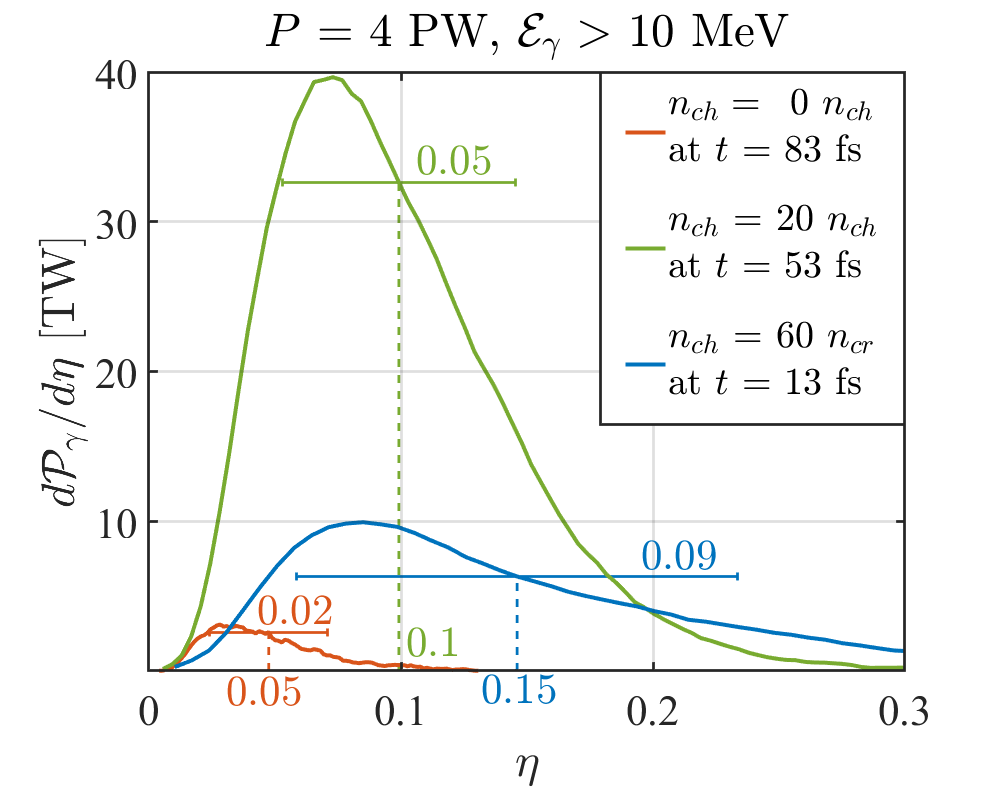}
      \caption{\label{fig:EtaDens} Normalized emitted power $\mathcal{P}_{\gamma}$ into a single lobe as a function of $\eta$ at the time of peak emission (see Fig.~\ref{fig:NphDens}) for $n_{ch} / n_{cr} = 0$, 20, and 60. The dashed lines show the average $\eta$ and the horizontal bars show the standard deviation.}
      \end{center}
\end{figure}

We continue this section by examining the key features of emission by individual electrons at $n_{ch} / n_{cr} = 0$ and $n_{ch} / n_{cr} = 60$. Figure~\ref{fig:EtaDens} shows the re-normalized emitted power $\mathcal{P}_{\gamma}$ [see Eq.~(\ref{P_norm})] as a function of $\eta$ for the three densities at the time of peak emission from Fig.~\ref{fig:NphDens}. As before, $\eta$ is a dimensionless parameter given by Eq.~(\ref{eta-AA}) that characterizes the electron acceleration in an instantaneous rest frame and determines the power of the synchrotron emission by the corresponding electron. The average $\eta$ at $n_{ch} / n_{cr} = 0$ is much lower than the average $\eta$ at the optimal density, which means that not only we have fewer emitting electrons in the initially empty channel, but that these electrons experience much weaker acceleration associated with the emission. However, the average $\eta$ at $n_{ch} / n_{cr} = 60$ has the highest value, indicating that individual electrons emit very efficiently in this regime. The reason for poor performance in terms of the conversion efficiency is that the emission of the electrons is not well-directed.

The parameter $\eta$ depends on the electron relativistic factor $\gamma_e$ and the effective field strength $\mathcal{F}_{eff}$ defined by Eq.~(\ref{F_eff_AA}). Figure~\ref{fig:FieldGammaDens} shows $\mathcal{P}_{\gamma}$ as a function of these two parameters in order to gain further insight into the photon emission process. It is evident that $\gamma_e$ and $\mathcal{F}_{eff}$ play different roles at $n_{ch} / n_{cr} = 0$ and  at $n_{ch} / n_{cr} = 60$. It is thus instructive to separately compare these two regimes to the regime with the optimal density.


\subsection{Photon emission at $n_{ch} / n_{cr} = 0$} \label{Sec-4A}

\begin{figure}
   \begin{center}
      \includegraphics[width=1\columnwidth]{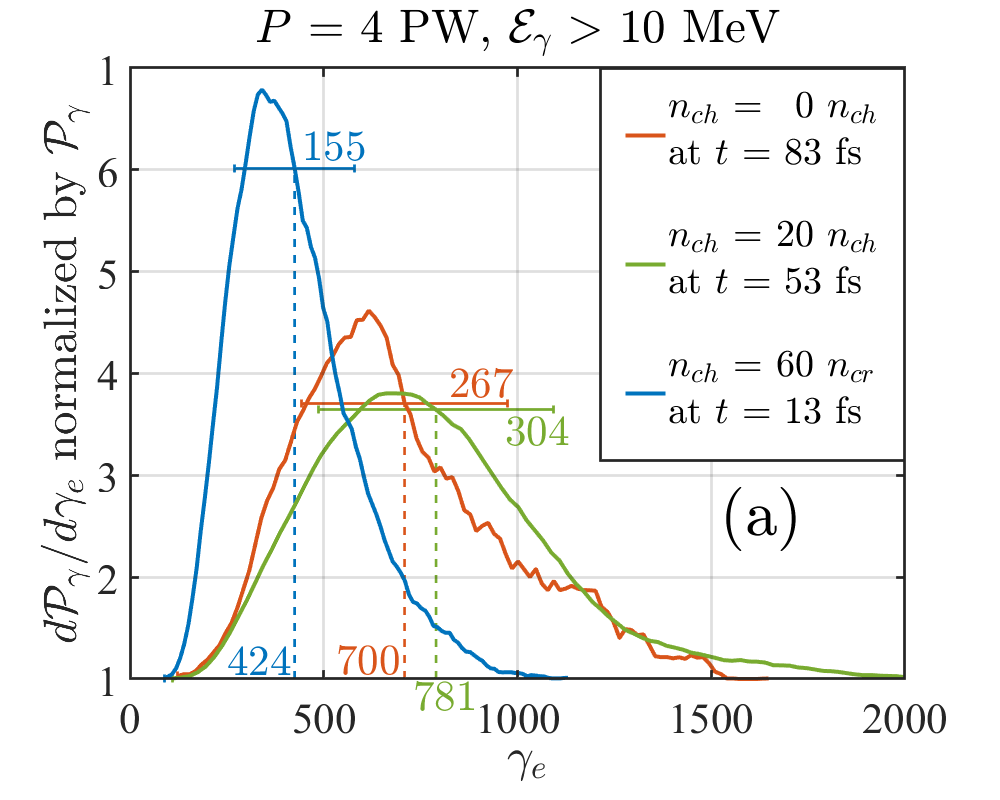}
      \includegraphics[width=1\columnwidth]{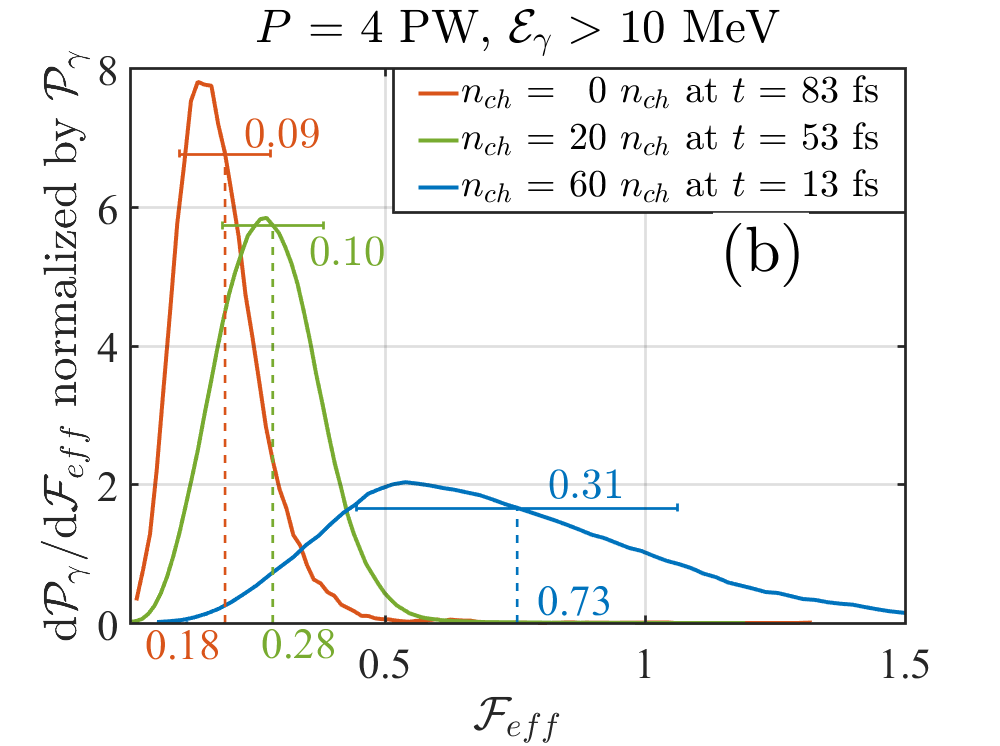}
      \caption{\label{fig:FieldGammaDens} Normalized emitted power $\mathcal{P}_{\gamma}$ as a function of $\mathcal{F}_{eff}$ and $\gamma_e$ at the time of peak emission (see Fig.~\ref{fig:NphDens}) for $n_{ch} / n_{cr} = 0$, 20, and 60. $\mathcal{P}_{\gamma}$ is for photons emitted into a single lobe. The dashed lines show the average $\mathcal{F}_{eff}$ and $\gamma_e$. The horizontal bars show the standard deviation.}
      \end{center}
\end{figure}

Figure~\ref{fig:FieldGammaDens}a confirms that the energy distribution of the emitting electrons in this case is very similar to that at the optimal density. Both the average $\gamma_e$ and the standard deviation are relatively close to those at $n_{ch} / n_{cr} = 20$.

The effective field strength $\mathcal{F}_{eff}$ however is considerably lower at $n_{ch} / n_{cr} = 0$ than at the optimal density. We found that the axial current density in an initially empty channel and the resulting azimuthal quasi-static magnetic field are much lower than at $n_{ch} / n_{cr} = 20$. The maximum strength of this magnetic field, $\bar{B}$, in the channel cross-section at the axial location of maximum emission only reaches $0.1 B_0$. We have already established that the number of electrons in an initially empty channel is reduced. The reduced quasi-static magnetic field is a manifestation of that.

The plasma magnetic field in this regime is so low that most of the emissions are caused by the laser fields. This is evident from the fact that the average value of $\mathcal{F}_{eff} \approx 0.2$ is two times higher than the maximum strength of the plasma field at the axial location of the emission, $\bar{B} / B_0 \approx 0.1$. We can further confirm the importance of the laser field by using simple estimates for $\eta$ in two limiting cases. If the plasma magnetic field dominates the emission, then $\eta$ is approximately given by
\begin{equation} \label{eta_B_AA}
    \eta \approx \gamma \bar{B}/E_S.
\end{equation}
If the laser fields dominate the emission, then we can estimate $\eta$ as 
\begin{equation} \label{eta_laser_AA}
    \eta \approx \gamma \left( 1 - \cos\theta \right) B_0/E_S.
\end{equation}
This estimate is obtained for an ultra-relativistic electron that is moving at an angle $\theta$ with respect to the wave-vector of a plane electromagnetic wave whose magnetic field amplitude is $B_0$. It follows from Eqs.~(\ref{eta_B_AA}) and (\ref{eta_laser_AA}) that the laser fields are more important than the plasma magnetic field if the angle $\theta$ satisfies the condition
\begin{equation} \label{theta_AA}
    \cos \theta < 1 - \bar{B} / B_0.
\end{equation}
According to Fig.~\ref{fig:HemispheresDens}a, $\theta \approx 30^{\circ}$ for the majority of the emissions. Then Eq.~(\ref{theta_AA}) is satisfied even for $\bar{B} / B_0 \approx 0.1$, which confirms that the laser fields are causing a significant part of the photon emission at $n_{ch} / n_{cr} = 0$.


\subsection{Photon emission at $n_{ch} / n_{cr} = 60$} \label{Sec-4B}

\begin{figure}
   \begin{center}
      \includegraphics[width=1\columnwidth]{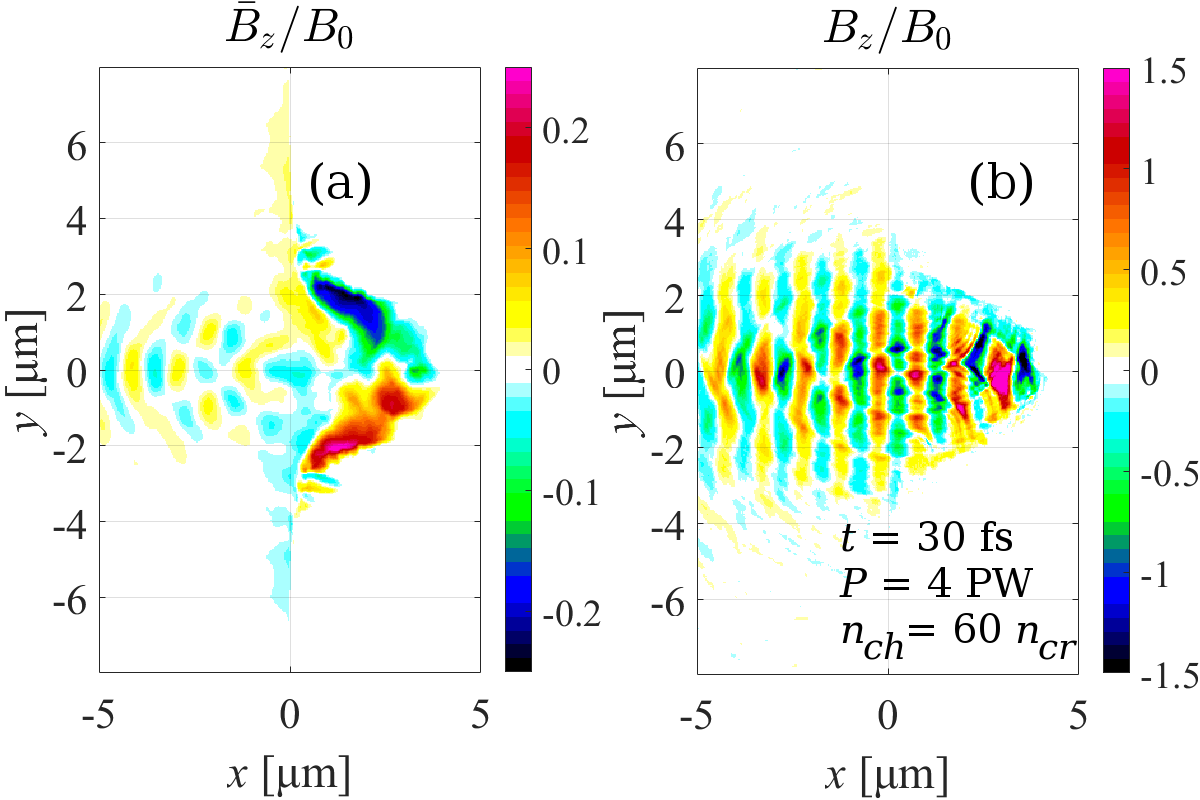}
      \caption{\label{fig:BfieldsDens} Time-averaged (a) and instantaneous (b) magnetic field for $n_{ch} = 60 n_{cr}$ at the time of peak emission. The laser propagates along the $x$-axis and its electric field is polarized along the $y$-axis. $B_0 \approx 2$~MT is the maximum magnetic field of the laser in the absence of the target. }
      \end{center}
\end{figure}

The enhancement in $\eta$ in this regime is caused by a significant enhancement in $\mathcal{F}_{eff}$ even though the electron spectrum is less energetic than at $n_{ch} / n_{cr} = 20$ and the characteristic $\gamma_e$ is also lower (see Fig.~\ref{fig:FieldGammaDens}a). Our analysis of the laser propagation shows that the laser pulse experiences a strong backward reflection at this channel density. As a result, instantaneous electric and magnetic fields across the channel are much higher than $E_0$ and $B_0$ even in the region where there is no strong beam focusing, as seen in Fig.~\ref{fig:BfieldsDens}b. On the other hand, the time-averaged over four laser periods magnetic field in Fig.~\ref{fig:BfieldsDens}a reaches only a maximum value of $0.3 B_0$. The reflected pulse creates a configuration where forward-moving electrons interact with a counter-propagating wave. It follows from Eq.~(\ref{F_eff_AA}) that $\mathcal{F}_{eff} \approx 2$ for a counter-propagating wave whose field amplitude is $E_0$. We thus conclude that the reflection is an underlying cause for the increase of $\mathcal{F}_{eff}$.

Even though $\eta$ is significantly higher at this density, the emitting electrons lack directivity. For example, even though the total number of electrons with the characteristic $\gamma_e$ is higher by a factor of two than that at the optimal density, the number of electrons whose momentum is confined within a $60^{\circ}$ cone is less, as seen in Fig.~\ref{fig:eHistoDens}b.

An important detail of our analysis should be reiterated: we have discussed the emission of photons into a narrow emission cone of only 5$^\circ$. Our motivation was to evaluate a performance of a gamma-ray source that can be used at larger distances from the laser-irradiated target. A much wider opening angle would lead to higher energy conversion rates. The conversion rate of laser energy into photons with $\mathcal{E}_{\gamma} > 1$~MeV that are emitted in the forward direction is 8.7$\%$ for $n_{ch} = 20 n_{cr}$, 4.5$\%$ for $n_{ch} = 60 n_{cr}$, and 1.0$\%$ for $n_{ch} = 0 n_{cr}$. 


\section{Creation of electron-positron pairs by colliding gamma-ray beams} \label{Sec-pair}

In Sec.~\ref{Sec-Den} we showed that the number of photons emitted into a narrow cone with an opening angle of $5^{\circ}$ increases rapidly with the incident laser power, provided that the channel has the optimal density. In fact, the number of multi-MeV photons increases as $P^2$ with the incident laser power $P$ while the laser intensity remains the same. Multiple applications can benefit from such a strong scaling. In this section, we examine one of them, which is generation of matter and antimatter directly from light via collisions of gamma-rays~\cite{PhysRevE.93.013201}. Other approaches do exist, such as the one where gamma-rays are fired into a the high-temperature radiation field of a laser-heated hohlraum ~\cite{Nature.8.434436}. 

\begin{figure} [H]
   \begin{center}
      \includegraphics[width=0.7\columnwidth,clip]{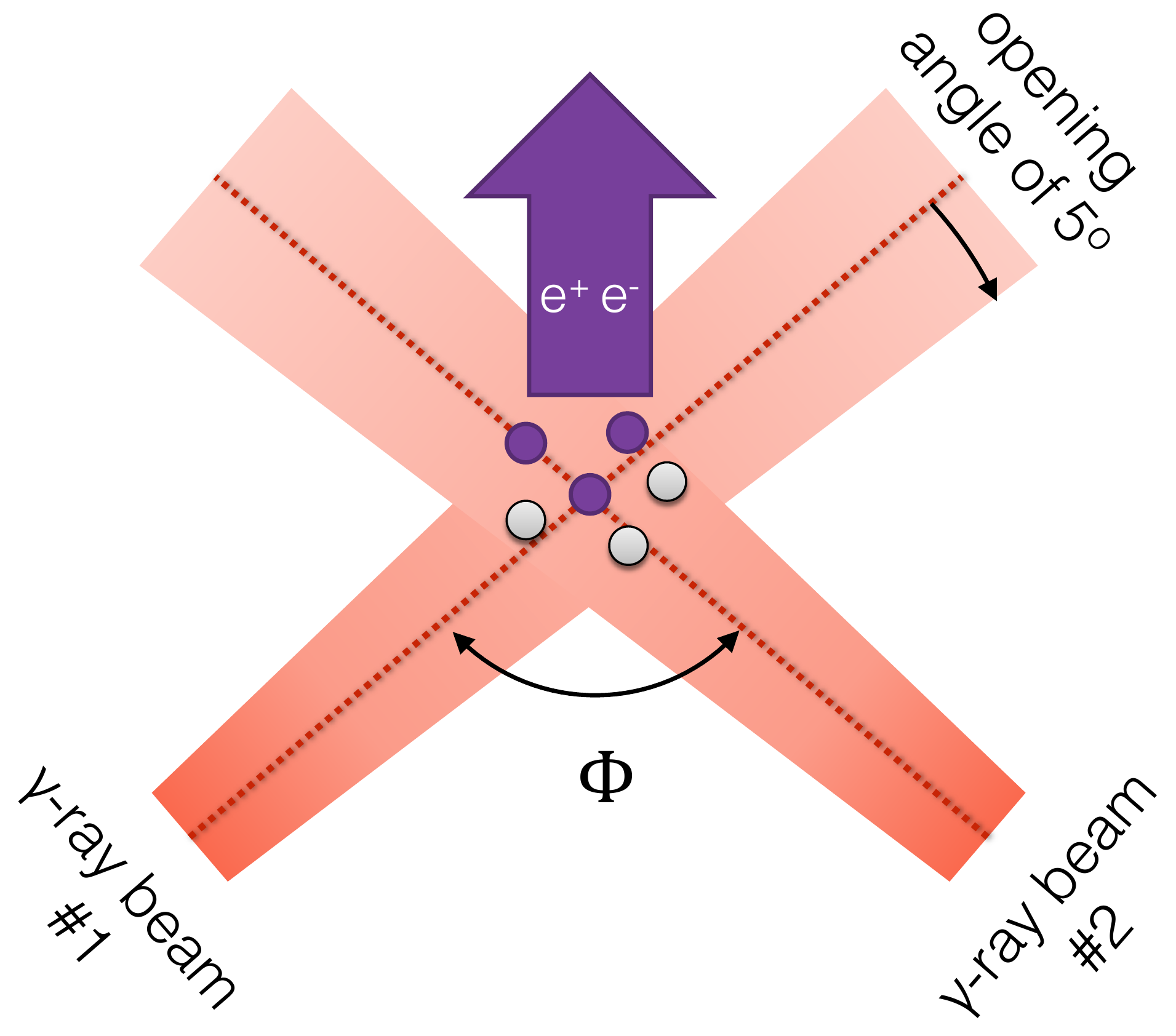}
      \caption{\label{fig:PosLasMomenta_90} Setup for pair production via two-photon collisions in two intersecting $\gamma$-ray beams.}
      \end{center}
\end{figure}

We are considering a setup that is schematically shown in Fig.~\ref{fig:PosLasMomenta_90} and that was first discussed in Ref.~\onlinecite{PhysRevE.93.013201}. The colliding gamma-ray beams are the collimated beams emitted into a single lobe and discussed in the preceding sections of the paper. The collision of the beams occurs in a vacuum at a distance $d$ that greatly exceeds the spatial scales associated with the photon emission within a single structured target. Multi-MeV photons are absolutely necessary for the pair production via collisions of two photons (linear Breit-Wheeler process~\cite{PhysRev.46.1087}) because the photon energies must satisfy the requirement $\mathcal{E}^a_{\gamma} \mathcal{E}^b_{\gamma} > 2 m_e^2 c^4$, where $\mathcal{E}^a_{\gamma}$ and $\mathcal{E}^b_{\gamma}$ are the energies of the colliding photons.

The increase of the multi-MeV photon numbers $N_{\gamma}$ produced by each of the targets significantly increases the number of produced pairs $N_{pairs}$. We first show this using simple estimates in order to derive the corresponding scaling. The photon density in each of the beams is $n_{\gamma} \approx N_{\gamma} / lS$, where $l$ is the beam length and $S$ is the beam cross-section at the location where the two beams collide. Taking into account that the collisions occur far from the emission region, we estimate that $S \approx \pi (\theta d)^2$, where $\theta = 5^{\circ}$ is the divergence angle of the photon beam. We assume that the two colliding beams have similar photon spectra. The number of pairs produced by one gamma-ray is roughly $\sigma_{\gamma \gamma} n_{\gamma} l$, where $\sigma_{\gamma \gamma}$ is the cross-section for the linear Breit-Wheeler process~\cite{PhysRev.46.1087}. The total number of pairs is $N_{\gamma}$ times higher. Taking into account the expressions for $n_{\gamma}$ and $S$, we find that
\begin{equation}
    N_{pairs} \approx N_{\gamma} \sigma_{\gamma \gamma} n_{\gamma} l \approx \sigma_{\gamma \gamma} N_{\gamma}^2 / \pi \theta^2 d^2 .
\end{equation}
The number of multi-MeV photons in our beams scales as $P^2$, which yields a strong scaling with the incident laser power $P$ for the number of electron-positron pairs:
\begin{equation} \label{pair_scaling}
    N_{pairs} \propto P^4 / d^2 .
\end{equation}

\begin{table}
\caption{Pair production via two-photon collisions at $90^{\circ}$}
\label{tab:Pairs}
\begin{tabular}{|l|c|c|c|}
\hline
 Incident power $P$&  &  & \\
 per laser beam & 1 PW & 2 PW & 4 PW \\
  \hline
  Number of multi-MeV &  &  & \\
  photons $N_{\gamma}$&  &  & \\
  per $\gamma$-ray beam & 1.5$\times 10^{11}$& 4.2$\times 10^{11}$& 2.8$\times 10^{12}$\\
  \hline
  Number of pairs $N_{pairs}$ & & & \\
  at $d = 500$~$\upmu$m & 15 & 140 & 3700\\
    \hline
  Number of pairs $N_{pairs}$ & & & \\
  at $d = 250$~$\upmu$m & 75 & 580 & $1.5 \times 10^4$\\
  \hline
  \end{tabular}
\end{table}

The derived scaling of $N_{pairs}$ with the laser power has been confirmed by calculating the corresponding reaction rate for the computed photon spectra as the ones shown in Fig.~\ref{fig:PhotEspectrumDens}b. The calculations were performed for a $\Phi = 90^{\circ}$ collision angle using the angle and energy dependent cross-section for the linear Breit-Wheeler process, as detailed in Ref.~\onlinecite{PhysRevE.93.013201}. We assume that the spectrum represents a uniform photon beam that diverges as a cone with a $5^{\circ}$ opening angle. Our results are summarized in Table~\ref{tab:Pairs} for $P = 1$, 2, and 4~PW. We have confirmed the number of pairs for the 1~PW case using a kinetic TrI-LEns simulation~\cite{JANSEN2018582} that directly checks for binary photon collisions. The number of pairs increases by a factor of 247 for $d = 250$ $\mu$m and by a factor of 200 for $d = 500$ $\mu$m as we increase the incident laser power by a factor of four. Equation~(\ref{pair_scaling}) predicts a factor of 256. As expected, the number of pairs also increases as $1/d^2$ when we reduce the distance between the gamma-ray sources and thus increases the density of the colliding photons.



Our photon spectra extend only down to $\mathcal{E}_{\gamma} = 1$~MeV, which means that the number of pairs produced by a photon spectrum extending to lower photon energies would be higher. However, calculating the spectra of photons emitted into a narrow cone becomes prohibitively difficult at lower energies due to a large total number of photons emitted in the sub-MeV energy range. We estimate that the error in $N_{pairs}$ associated with the sub-MeV photons is roughly 25\%. We obtained this number by extrapolating our spectra down to $\mathcal{E}_{\gamma} = 0.1$~MeV.

\section{Summary and Discussion} \label{Sec-conc}

We have examined the emission of collimated gamma-ray beams from structured laser-irradiated targets with pre-filled and empty cylindrical channels. Our goal was to determine how the conversion efficiency of the laser energy into gamma-rays scales with the incident laser power when the peak intensity is held constant. We found that the conversion efficiency into a narrow beam of gamma-rays with a $5^{\circ}$ opening angle can be significantly increased by utilizing channels with an optimal density. The conversion efficiency into multi-MeV photons increases roughly linearly with the incident laser power in the range between 1 and 4 PW. The optimal range of densities for the peak laser intensity of $5 \times 10^{22}$ W/cm$^2$ is between 10 and $20 n_{cr}$. The considered setup involving a structured target becomes a powerful gamma-ray source when irradiated by a 4~PW laser pulse. At the optimal target density, the power emitted by multi-MeV photons into a $5^{\circ}$ cone reaches 143~TW. The resulting gamma-ray intensity 1 cm away from the emission location is $6 \times 10^{15}$~W/cm$^2$. 

Our detailed particle tracking has revealed that the underlying cause for the improved conversion efficiency is the enhanced electron acceleration in the laser-pulse assisted by the quasi-static magnetic field generated in the channel~\cite{gong2018forward}. This is a nontrivial result, because the laser intensity and thus $a_0$ in the incoming laser pulse remain unchanged. Another peculiar aspect revealed by our analysis is that the characteristic effective field $\mathcal{F}_{eff}$ experienced by the emitting electrons in the channel undergoes only a modest increase even though the azimuthal magnetic field becomes stronger by almost a factor of two as the channel becomes wider. It is likely that the improved transverse electron confinement by the magnetic field plays a significant role at higher power $P$, as the channel radius is increased as $\sqrt{P}$ in our scan. The changes in the longitudinal laser-driven electron current might also be important, since the acceleration mechanism directly depends on the current density~\cite{gong2018forward}. A detailed study that also involves electron tracking along their trajectories is required to pinpoint the exact changes causing the observed energy increase. 

We have also examined how the increased conversion efficiency impacts the yield of electron-positron pairs in a collision of two gamma-ray beams. We found that the number of generated pairs increases as $P^4$ at a fixed laser intensity as opposed to just $P^2$, which one would expect in the case of a constant conversion efficiency. Our results indicate that increasing the peak laser intensity is not necessary to achieve a dramatic increase in the number of generated electron-positron pairs.

This work also emphasizes the need for the target development since achieving the observed power scaling requires pre-filled targets. Empty structured targets have already been used experimentally to achieve greater control over laser interactions with solid-density targets~\cite{snyder2019relativistic}. Advanced target manufacturing facilities are also able to produce solid targets of variable density in the desired range of 10 to $20 n_{cr}$ (when fully ionized and homogenized) using the in situ polymerization technique. The pore and thread structures are sub-micron, so a relatively homogeneous plasma has been achieved in experiments with high-intensity lasers~\cite{Willingale_2018}. Currently, the challenge is to manufacture pre-filled channels that are several microns in radius. In that context, higher power experiments would be more feasible since they require wider channels that are easier to fill. 

Finally, it must be stated that our results for the initially empty channels qualitatively differ from some of the results that have been published over the last several years~\cite{PhysRevLett.116.115001,yu2018generation,yu_2019}. It has been reported that empty channels produce highly collimated gamma-ray beams with just a single lobe aligned with the direction of the laser propagation. This emission pattern is achieved via longitudinal electron acceleration without transverse oscillations across the channel. In a pre-filled channel, both quasi-static electric and magnetic fields induce a force causing electrons to move towards the axis and, as a result, oscillate. In an empty channel, the electric field can be directed radially inward, causing a force compensation and preventing transverse oscillations~\cite{Gong_hollow_channel}. However, inward ion motion disrupts the force balance, leading to electron oscillations, which is exactly what happens in our simulations. The importance of the ion motion increases with laser intensity~\cite{Wang_2019}. The ion physics in the simulations with highly collimated photon beams was suppressed by: 1) using an extremely short laser pulse with an extremely sharp rising edge and 2) using heavy ions with a prescribed relatively low ionization state. The extreme sensitivity of the electron and photon collimation to the temporal profile of the laser pulse in initially empty channels means that quantitative predictions require simulations with an experimentally measured temporal profile. This also indicates that achieving experimentally high efficiency and high collimation in initially empty channels might be extremely challenging. In contrast to that, pre-filled structured targets offer a robust setup for efficient generation of gamma-ray beams.  


\section*{Acknowledgements}
This research was supported by NSF (Grants No. 1632777 and 1821944) and AFOSR (Grant No. FA9550-17-1-0382). Simulations were performed with EPOCH (developed under UK EPSRC Grants No. EP/G054940/1, No. EP/G055165/1, and No. EP/G056803/1) using HPC resources provided by TACC at the University of Texas. This work used XSEDE, supported by NSF grant number ACI-1548562. Data collaboration supported by the SeedMe2 project (http://dibbs.seedme.org).

\section*{References}
\input{output.bbl}


\appendix

\section{Magnetic field and photon emission profiles in the power scan} \label{Appendix_B}

\begin{figure*}
      \centering
      \includegraphics[width=1\textwidth,trim={0.0cm 0.0cm 0.0cm 0.0cm},clip]{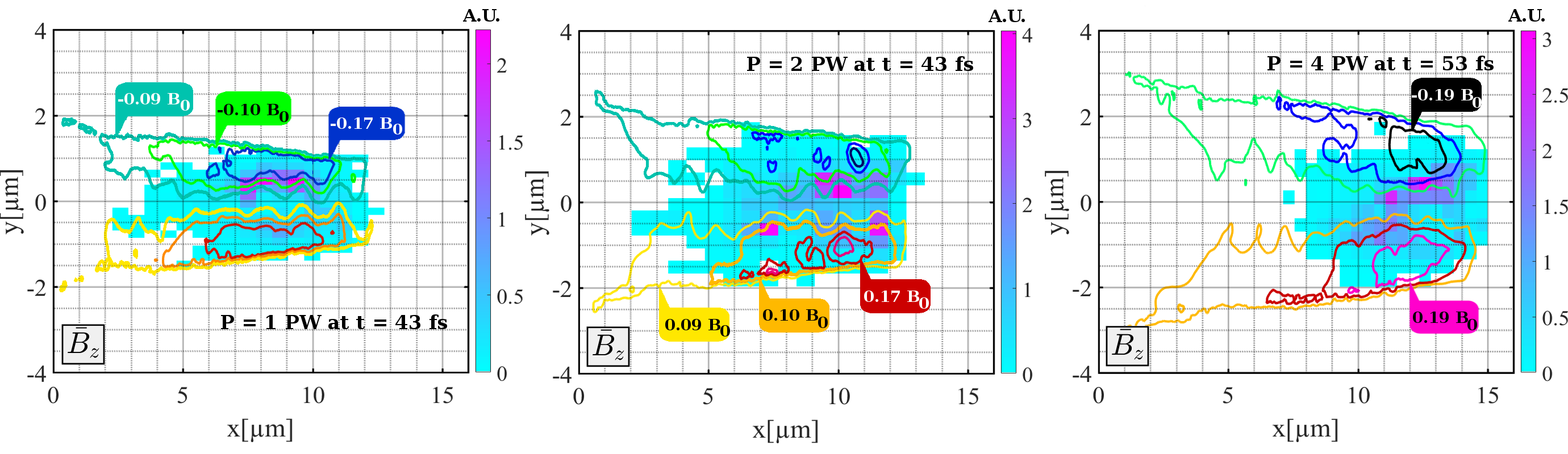}
      \caption{\label{fig:PosPh} Quasi-static magnetic field, $\bar{B}_z$, and energy density of the photon emission with $\mathcal{E}_{\gamma} > 10$~MeV. Here $B_0 \approx 2$~MT is the maximum magnetic field of the laser pulse in the abscence of the target.}
\end{figure*}

Figure~\ref{fig:PosPh} shows profiles of a time-averaged azimuthal magnetic field and photon emission for $P = 1$, 2, and 4~PW in the power scan of Sec.~\ref{Sec-Pow}. The magnetic field is obtained by averaging the instantaneous magnetic fields over four laser periods. The time averaging is indicated using the overline. The contours show the magnetic field, $\bar{B}_z$, in the $(x,y)$-plane at $z = 0$ at the time of maximum emission from Fig.~\ref{fig:IphOvert}. The color in Fig.~\ref{fig:PosPh} shows the energy density associated with the emission of photons with $\mathcal{E}_{\gamma} > 10$~MeV. We only count the photons emitted in a slab with $|z| < 1$ $\mu$m over a 10 fs window around the time of maximum emission from Fig.~\ref{fig:IphOvert}. 

\onecolumngrid 

\end{document}

%% file: output.bbl
%